\documentclass[preprints,article,accept,moreauthors,pdftex]{mdpi} 

\firstpage{1} 
\pubvolume{xx}
\issuenum{1}
\articlenumber{5}
\pubyear{2019}
\copyrightyear{2019}
\history{Received: date; Accepted: date; Published: date}

\pdfoutput=1

\Title{Estimating Time-Varying Applied Current in the Hodgkin-Huxley Model}

\Author{Kayleigh Campbell, Laura Staugler and Andrea Arnold*\orcidA{}}

\AuthorNames{Kayleigh Campbell, Laura Staugler and Andrea Arnold}

\address[1]{Department of Mathematical Sciences,
  Worcester Polytechnic Institute,
  Worcester, MA 01609, USA; 
  kacampbell@wpi.edu (K.C.); lstaugler@wpi.edu (L.S.); anarnold@wpi.edu (A.A.)}

\corres{Correspondence: anarnold@wpi.edu; Tel.: +1-508-831-6825}

\abstract{The classic Hodgkin-Huxley model is widely used for understanding the electrophysiological dynamics of a single neuron.  While applying a constant current to the system results in a single voltage spike, it is possible to produce more interesting dynamics by applying time-varying currents, which may not be experimentally measurable.  The aim of this work is to estimate time-varying applied currents of different deterministic forms given noisy voltage data.  In particular, we utilize an augmented ensemble Kalman filter with parameter tracking to estimate four different deterministic applied currents, analyzing how the model dynamics change in each case.  We test the efficiency of the parameter tracking algorithm in this setting by exploring the effects of changing the standard deviation of the parameter drift and the frequency of data available on the resulting time-varying applied current estimates and related uncertainty.}


\begin{document}



\section{Introduction}

The Hodgkin-Huxley model is a classical system of differential equations that is widely-used for understanding the electrophysiological dynamics of a single neuron \cite{HodHux1952}.  The model is based on a simple circuit analogy, where each piece of the circuit corresponds to an electrophysiological component, representing the resistance of an electrically charged ion channel as a function of time and voltage \cite{Ermentrout2010, Nelson2004}.  While the Hodgkin-Huxley equations can be used to model the total current resulting from an applied voltage, the model can also be used to predict voltage given an externally applied current.  The latter is particularly useful in experimental settings where voltage measurements are obtainable, making it possible to estimate the applied current based on observed voltage data.  

In the standard model formulation, applying a constant current to the system results in a single voltage spike.  However, by applying time-varying currents, it is possible to produce more interesting dynamics that include multiple voltage spikes.  The work in this paper focuses on how the Hodgkin-Huxley system dynamics change when various deterministic, time-varying currents are applied.  More specifically, the aim of this work is to estimate time-varying applied currents of different deterministic functional forms given some noisy observations of voltage.  To tackle this inverse problem, we utilize an augmented ensemble Kalman filter (EnKF) with parameter tracking to estimate the Hodgkin-Huxley model states and time-varying applied current parameter.  

Various methods have been used in the literature to estimate certain constant (or static) parameters in the Hodgkin-Huxley equations; see, e.g., \cite{Lankarany2014, Vavoulis2012, Buhry2012, Doi2002}.  However, the focus of this work is on estimating the time-varying applied current parameter, which we assume is unmeasurable with unknown dynamics.  While many parameter estimation methods are available in the literature, the EnKF is particularly useful for the problem at hand due to the sequential nature of the algorithm's updating scheme, which corrects the model prediction with the available data one point at a time \cite{Burgers1998, Evensen1994}.  If the time-varying parameter changes more slowly than the system dynamics, it is possible to track the change in the parameter over time using a random walk \cite{Voss2004, Hamilton2013, Arnold2019}.  Further, since unknowns are treated as random variables in the Bayesian framework, there is a natural measure of uncertainty in the resulting parameter estimates, which lies in the estimated ensemble covariances of the underlying posterior probability distributions \cite{Kaipio2005, Calvetti2007}.

We analyze this problem using synthetic voltage data generated by applying four different deterministic functions as the applied current -- a constant current, a step function with one long step, a step function with multiple shorter steps, and a sinusoidal function -- in order to track and observe how the model dynamics change in each of these cases.  We further test the efficiency of the EnKF with parameter tracking algorithm in these cases by performing numerical experiments first to establish baseline results, then to explore the effects of changing the standard deviation of the drift term in the parameter tracking algorithm as well as the frequency of data available on the resulting applied current parameter estimates. 

The paper is organized as follows.  Section 2 gives a brief review of the Hodgkin-Huxley model, summarizing the relevant equations.  Section 3 describes the parameter estimation inverse problem and outlines the ensemble Kalman filtering algorithm, describing in particular time-varying parameter estimation using the EnKF with parameter tracking.  Section 4 gives the numerical results, including the generation of synthetic data and the numerical experiments relating to estimating the time-varying applied current parameter, and Section 5 features a discussion of the results and future work.


\section{Review of the Hodgkin-Huxley Model Equations}

\begin{figure}[t!]
\centering{\includegraphics[width=0.58\textwidth]{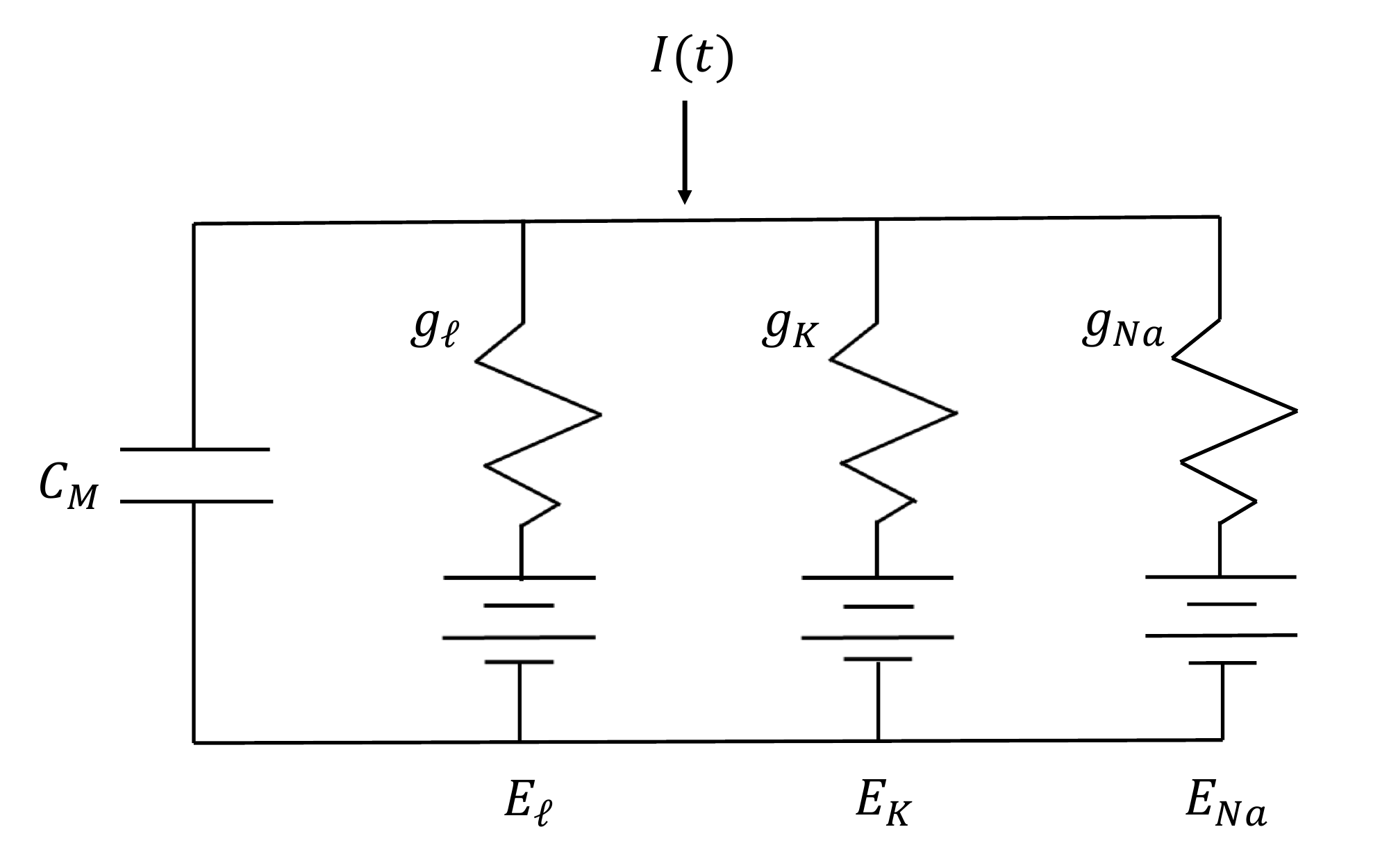}}
\caption{The Hodgkin-Huxley model represented as a circuit.  Here the capacitor ($C_M$) represents the charge storage capacity, the resistors ($g_{\ell}$, $g_K$, and $g_{Na}$) act as the ion channels, and the batteries ($V_{\ell}$, $V_K$, and $V_{Na}$) act as the electrochemical potentials.}
\label{Fig:circuit}
\end{figure}

The Hodgkin-Huxley model provided the first quantitative description of electrical excitability in nerve cells \cite{Schwiening2012}, involving detailed mathematical equations to describe the voltage-dependent and time-dependent properties of the sodium and potassium conductances \cite{HodHux1952}.  Each piece of the circuit shown in Figure \ref{Fig:circuit} corresponds to a different electrophysiological component of the model.  Capacitors represent the charge storage capacity of each gating variable; resistors represent the sodium, potassium, and leakage ion channels in the neuron; and batteries represent the electrochemical potentials that each gating variable has to let ions in and out of the charged cell.  

From this analogy, the Hodgkin-Huxley equation modeling total membrane current is given by
\begin{equation}\label{Eq:Tot_Mem_Current}
I = C_{M}\frac{dV}{dt} + I_{ion}
\end{equation}
where $I = I(t)$ is the total membrane current in the axon (with a positive inward current), $C_M$ is the membrane capacity (assumed to be constant), and $V = V(t)$ is the displacement of the membrane potential from its resting value (assumed to have a negative depolarization).  Table~\ref{tab:HH_eq} lists each model component, along with its corresponding units.  For simplicity of terminology, we will refer interchangeably to $V(t)$ as the voltage within this paper.  Note that the voltage $V$ is related to the membrane potential $E$ via the relationship $V = E - E_{r}$, where $E_{r}$ denotes the absolute value of the resting potential \cite{HodHux1952}.

\begin{table}[t!]
\centering
\begin{tabular}{|l|l|l|l|}
\hline
\textbf{Component} & \textbf{Description} & \textbf{Units} \\ \hline
$I$ & Total membrane current & mA/cm\textsuperscript{2} \\ \hline
$C_{M}$ & Membrane capacity & $\mu$F/cm\textsuperscript{2} \\ \hline
$V$ & Voltage & mV \\ \hline
$I_{ion}$ & Ionic current density & mA/cm\textsuperscript{2} \\ \hline
$t$ & Time & msec \\ \hline
\end{tabular}
\caption{Components of the Hodgkin-Huxley model \eqref{Eq:Tot_Mem_Current}.}
\label{tab:HH_eq}
\end{table}

The ionic current density $I_{ion}$ is represented as the sum of the three currents
\begin{equation}\label{Ionic_current}
I_{ion} = I_{Na} + I_{K} + I_{\ell}
\end{equation}
where $I_{Na}$, $I_K$, and $I_{\ell}$ model the currents relating to the sodium, potassium, and leakage channels occurring in the neuron, respectively.  The form of the current for each ion channel follows from Ohm's law, where
\begin{equation}
I_i = g_i(t,V)(V-V_i)
\end{equation}
for $i = Na, K, \ell$.  Here $g_i(t,V)$ represents the gate for each channel generally as a function of time and voltage, and $V-V_i$ represents the difference between the overall voltage $V$ of the system and the channel-specific voltages $V_i$.  We describe each of the three ionic currents in more detail as follows.

\textbf{\textit{Sodium current.}}  The sodium current is given by
\begin{equation}\label{Eq:Na_current}
I_{Na} = g_{Na}(t,V)(V - V_{Na})
\end{equation}
where the sodium gate
\begin{equation}\label{Eq:Na_conductance}
g_{Na}(t,V) = m^{3}h\bar{g}_{Na}
\end{equation}
is impacted by depolarization, which causes an increase in sodium conductance \cite{HodHux1952}.  Here $\bar{g}_{Na}$ is a constant (conductance/cm\textsuperscript{2}), $m = m(t)$ is the proportion of active sodium gates open (dimensionless variable which varies over time between 0 and 1), and $h = h(t)$ is the proportion of inactive gates open (similarly dimensionless, varying between 0 and 1).  The sodium voltage is given by $V_{Na} = E_{Na} - E_{r}$, where $E_{Na}$ is an equilibrium potential for sodium.  Table~\ref{tab:constants} lists the constant values of $V_{Na}$ and $\bar{g}_{Na}$.

The dynamics of the sodium gating variables $m(t)$ and $h(t)$ are governed by the following differential equations:
\begin{eqnarray}
\frac{dm}{dt} &=& \alpha_{m}(V)(1 - m) - \beta_{m}(V)m \label{Eq:m_transfer} \\
\noalign{\vspace{.1cm}}
\frac{dh}{dt} &=& \alpha_{h}(V)(1 - h) - \beta_{h}(V)h \label{Eq:h_transfer}
\end{eqnarray}
where the voltage-dependent rate constants (msec\textsuperscript{-1}) $\alpha_m$ and $\alpha_h$ represent the rate of flow of ions into the cell and $\beta_m$ and $\beta_h$ represent the flow out.  The rate constants are modeled using the following equations, derived from Hodgkin and Huxley's experimental results \cite{HodHux1952}:
\begin{eqnarray}
\alpha_{m}(V) &=& \frac{0.1(V + 25)}{\exp(\frac{V + 25}{10}) - 1} \label{Eq:m_alpha} \\
\noalign{\vspace{.1cm}}
\alpha_{h}(V) &=& 0.07\exp\Big(\frac{V}{20}\Big) \label{Eq:h_alpha} \\
\noalign{\vspace{.1cm}}
\beta_{m}(V) &=& 4\exp\Big(\frac{V}{18}\Big) \label{Eq:m_beta} \\
\noalign{\vspace{.1cm}}
\beta_{h}(V) &=& \frac{1}{\exp(\frac{V +30}{10}) + 1} \label{Eq:h_beta}
\end{eqnarray}
Note that setting $\alpha_{m}$ to its limit value of 1 at $V = -25$ mV avoids the discontinuity at that point.

\textbf{\textit{Potassium current.}}  The potassium current is given by
\begin{equation}\label{K_current}
I_{K} = g_{K}(t,V)(V - V_{K})
\end{equation}
with potassium gate equation
\begin{equation}\label{Eq:K_conductance}
g_{K}(t,V) = \bar{g}_{K}n^{4}
\end{equation}
based on the assumption that potassium ions must have four similar particles in order to cross the membrane.  Here $\bar{g}_{K}$ is a constant (conductance/cm\textsuperscript{2}) and $n = n(t)$ is the proportion of potassium gates open (dimensionless, varying between 0 and 1).  The potassium voltage is given by $V_K = E_{K} - E_{r}$, where $E_{K}$ is an equilibrium potential for potassium, sensitive to the overall outside concentration of charged ions \cite{HodHux1952b}.  Table~\ref{tab:constants} lists the constant values of $V_{K}$ and $\bar{g}_{K}$.

The dynamics of the gating variable $n(t)$ are similarly modeled using the differential equation
\begin{equation}\label{Eq:n_transfer}
\frac{dn}{dt} = \alpha_{n}(V)(1 - n) - \beta_{n}(V)n
\end{equation}
where the rate constants 
\begin{eqnarray}
\alpha_{n}(V) = \frac{0.01(V + 10)}{\exp(\frac{V + 10}{10}) - 1} \label{Eq:n_alpha} \\
\noalign{\vspace{.1cm}}
\beta_{n}(V) = 0.125\exp\Big(\frac{V}{80}\Big) \label{Eq:n_beta}
\end{eqnarray}
were also derived using experimental data \cite{HodHux1952}.  Note the discontinuity in $\alpha_n$ when $V = -10$ mV can be avoided by setting it equal to its limit value of 0.1 at that point.

\textbf{\textit{Leakage current.}}  The leakage current is a small combined current, accounting mostly for chloride but also other ions.  The leakage current is given by
\begin{equation}\label{leak_current}
I_{\ell} = \bar{g}_{\ell}(V - V_{\ell})
\end{equation}
with constant conductance $\bar{g}_\ell$ and leakage voltage $V_{\ell} = E_{\ell} - E_{r}$.  Here $E_{\ell}$ is the potential at which the leak current is zero.  The leakage voltage $V_\ell$ is needed for any calculation for threshold, but it is unlikely to give any information about the nature of charged particles \cite{HodHux1952b}.  Table~\ref{tab:constants} lists the constant values of $V_\ell$ and $\bar{g}_{\ell}$.

\begin{table}[t!]
\centering
\begin{tabular}{|l|l|l|l|}
\hline
\textbf{Parameter} & \textbf{Description} & \textbf{Value} & \textbf{Units}\\ \hline
$C_{M}$ & Membrane capacity & 1.0 & $\mu$F/cm\textsuperscript{2} \\ \hline
$V_{Na}$ & Sodium voltage & -115 & mV \\ \hline
$V_{K}$ & Potassium voltage & 12  & mV \\ \hline
$V_{\ell}$ & Leakage voltage & -10.613 & mV \\ \hline
$\bar{g}_{Na}$ & Sodium gate constant & 120 & m.mho/cm\textsuperscript{2} \\ \hline
$\bar{g}_{K}$ & Potassium gate constant & 36 & m.mho/cm\textsuperscript{2} \\ \hline
$\bar{g}_{\ell}$ & Leakage gate constant & 0.3 & m.mho/cm\textsuperscript{2} \\ \hline
\end{tabular}
\caption{Constant parameter values used in the Hodgkin-Huxley model. \textit{Note that m.mho stands for 1/ohm (the reverse of ohm) or amp/volts. It is the unit Siemen and represents the derived unit of electrical conductance.}}
\label{tab:constants}
\end{table}

\textbf{\textit{Model summary.}}  In summary, the Hodgkin-Huxley model comprises the total membrane current equation \eqref{Eq:Tot_Mem_Current}, which depends on time, voltage, and the solutions to the transfer equations \eqref{Eq:m_transfer}, \eqref{Eq:h_transfer}, and \eqref{Eq:n_transfer}.  Note that when a constant voltage is applied, $\frac{dV}{dt}=0$ and \eqref{Eq:Tot_Mem_Current} simplifies to $I = I_{ion}$.  In this case, equations \eqref{Eq:m_transfer}, \eqref{Eq:h_transfer}, and \eqref{Eq:n_transfer} can be solved independently to compute the total ionic current.  

However, when voltage changes with time due to an applied current, $\frac{dV}{dt}\neq0$ and all four equations must be solved simultaneously.  The complete system of coupled ordinary differential equations is given by
\begin{eqnarray}
\frac{dV}{dt} &=& \frac{1}{C_M}(I - I_{ion})  \label{Eq:HH1} \\
\noalign{\vspace{.1cm}}
\frac{dn}{dt} &=& \alpha_{n}(V)(1-n) - \beta_{n}(V)n \label{Eq:HH2} \\
\noalign{\vspace{.1cm}}
\frac{dm}{dt} &=& \alpha_{m}(V)(1-m) - \beta_{m}(V)m  \label{Eq:HH3} \\
\noalign{\vspace{.1cm}}
\frac{dh}{dt} &=& \alpha_{h}(V)(1-h) - \beta_{h}(V)h  \label{Eq:HH4}
\end{eqnarray}
Note that in this case, the applied current $I = I(t)$ in \eqref{Eq:HH1} drives the system dynamics.  We explore how different choices of deterministic, time-varying functions for $I(t)$ affect the dynamics of the system in the numerical results.


\section{Parameter Estimation and the Ensemble Kalman Filter}
\label{Sec:ParamEst}

Given measurements of voltage, our aim is to estimate the time-varying applied current $I(t)$ that best fits the available data.  This is a parameter estimation inverse problem, where the parameter of interest is a time-varying deterministic function with assumed unknown dynamics.  More specifically, we assume here that we cannot directly measure the time-varying applied current and that we do not have equations available to explain its dynamics. 

The set-up for this inverse problem is similar to the standard set-up for estimating parameters in initial value problems of the form 
\begin{equation}\label{Eq:dx}
\frac{dx}{dt} = f(t,x,\theta), \qquad x(0) = x_0
\end{equation}
where $x = x(t)$ denotes the model states and $\theta$ denotes the model parameters \cite{Arnold2014}.  Given some discrete, noisy system measurements
\begin{equation}\label{Eq:obs}
y_j = G(x(t_j), \theta) + w_j, \qquad 0< t_1 < \dots < t_T
\end{equation}
the inverse problem is to estimate the model states $x(t)$ and parameters $\theta$.  Most classical approaches addressing this problem tend to focus on the case when the parameters are constants, i.e., when $\frac{d\theta}{dt} = 0$.  In this case, however, $\theta=\theta(t)$ and $\frac{d\theta}{dt}$ is some unknown function.  

To estimate the time-varying applied current in this work, we use a version of the ensemble Kalman filter (EnKF) with parameter tracking \cite{Arnold2014, Arnold2019}.  The EnKF is an extension of the classical Kalman filter adapted to work with models that are not necessarily linear or Gaussian \cite{Burgers1998, Evensen1994}.  As a Bayesian statistical algorithm, the EnKF treats all unknowns as random variables with corresponding probability distributions.  The filter uses a random sample to represent the current probability distribution of states and parameters, then utilizes ensemble statistics along with model predictions and observed data to update the sample at each discrete time point.  

While the original EnKF was implemented for state estimation, the augmented EnKF allows for simultaneous state and parameter estimation \cite{Evensen2009}.  The steps of the augmented EnKF are summarized as follows.  At time $j$, the sample 
\begin{equation}
S_{j \mid j} = \Big\{(x_{j \mid j}^1,\theta_{j \mid j}^1),(x_{j \mid j}^2,\theta_{j \mid j}^2), \dots , (x_{j \mid j}^N,\theta_{j \mid j}^N)\Big\}
\end{equation}
gives a discrete representation of the probability distribution, which is then updated using a two-step process to time $j+1$.  In the first step (the prediction step), we solve the system \eqref{Eq:dx} to predict the state values at time $j+1$.  In this work, \eqref{Eq:dx} is the Hodgkin-Huxley model given in \eqref{Eq:HH1}--\eqref{Eq:HH4}.  The state prediction ensemble is computed using the equation
\begin{equation}
x_{j+1 \mid j}^p = F(x_{j \mid j}^p, \theta_{j}^p) + v_{j+1}^p, \qquad p = 1,2,\dots,N
\end{equation}
where $F$ is the solution to \eqref{Eq:dx} at time $j+1$ and $v_{j+1}^p\sim\mathcal{N}(0,\mathsf{C})$ represents error in the model prediction.  The predicted states $x_{j+1\mid j}^p$ and current parameter values $\theta_j^p$ are then placed in the augmented vectors
\begin{equation}
z_{j+1\mid j}^p = \left[ \begin{array}{c} x_{j+1\mid j}^p \\ \theta_j^p \end{array} \right], \qquad p = 1,2,\dots,N
\end{equation}
which are used to compute ensemble statistics.  The prediction ensemble mean is computed using the formula 
\begin{equation}
\bar{z}_{j+1 \mid j} = \frac{1}{N}\sum_{p=1}^N z_{j+1 \mid j}^p
\end{equation}
and the prior covariance matrix by
\begin{equation}
\Gamma_{j+1 \mid j} = \frac{1}{N-1}\sum_{p=1}^N (z_{j+1 \mid j}^p - \bar{z}_{j+1 \mid j})(z_{j+1 \mid j}^p - \bar{z}_{j+1 \mid j})^T.
\end{equation}
Note that while the parameters $\theta_j^p$ are not updated in the prediction step, their cross-correlation information with the predicted states is embedded in the prior covariance matrix and is used in the next step to update the posterior sample.

In the second step (the observation update), the predicted values are compared with the observed data $y_{j+1}$ at time $j+1$.  The observation ensemble
\begin{equation}
y_{j+1}^p = y_{j+1} + w_{j+1}^p, \qquad p = 1,2,\dots,N
\end{equation}
where $w_{j+1}^p\sim\mathcal{N}(0,\mathsf{D})$ represents the observation error, is compared to the observation model predictions
\begin{equation}
\hat{y}_{j+1}^p = G(x_{j+1 \mid j}^p, \theta_{j}^p), \qquad p = 1,2,\dots,N
\end{equation}
computed using the observation model $G$ as in \eqref{Eq:obs}.  In this work, $G$ is a linear observation function measuring only the voltage in the Hodgkin-Huxley system.  The combined posterior ensemble is then given by
\begin{equation}
z_{j+1 \mid j+1}^p = z_{j+1 | j}^p + K_{j+1}(y_{j+1}^p - \hat{y}_{j+1}^p)
\end{equation}
for each $p = 1,2,\dots,N$.  The Kalman gain matrix $K_{j+1}$ incorporates the cross-covariance of state and model predications, the forecast error covariance, and the observation noise covariance.  For additional implementation details, see \cite{Arnold2018, Arnold2014}.

In the above formulation of the augmented EnKF, $\theta$ is assumed to be constant and is evolved artificially with time in order to obtain an estimate.  When $\theta$ is time-varying, as in the case we are considering, the augmented EnKF as presented requires additional modification.  If $\theta=\theta(t)$ changes more slowly than the dynamics of the system, it is possible to track the changes in $\theta(t)$ by incorporating a random walk in the prediction step.  To implement this, a new random variable $\xi$ is introduced and added to current estimate of $\theta$ at each prediction step, allowing the previously fixed parameters to take a random walk of the form
\begin{equation}\label{Eq:rand_walk}
\theta_{j+1 \mid j}^p = \theta_{j \mid j}^p + \xi_{j}^p, \qquad \xi_{j}^p \sim\mathcal{N}(0,\sigma_{\xi}^{2})
\end{equation}
for each $p = 1,2,\dots,N$.  Parameter tracking of this type has been used in various data assimilation problems; see, e.g., \cite{Voss2004, Hamilton2013, Arnold2019, Matzuka2014}.  Here we note that the choice of the standard deviation $\sigma_{\xi}$ in the drift term of the random walk is important in the accuracy and uncertainty of the resulting parameter estimate.  We will explore this further in the numerical experiments.


\section{Numerical Experiments}

In this section we describe the numerical experiments performed using the augmented EnKF with parameter tracking to estimate the time-varying applied current in the Hodgkin-Huxley model.  All experiments were performed using the MATLAB\textsuperscript{\textregistered} programming language.  In particular, we used the built-in solver \verb"ode15s" to numerically solve \eqref{Eq:HH1}--\eqref{Eq:HH4} due to the potential stiffness of the system when applying different time-varying currents.  We first describe how synthetic data was generated using four deterministic applied currents.  We then provide the numerical results obtained using parameter tracking to estimate the applied current in each case, testing the effects of changing the standard deviation of the parameter drift in \eqref{Eq:rand_walk} and the amount of available data.

\subsection{Synthetic Data Generation}
\label{SubSec:Data}

\begin{figure*}[h!]
\centerline{\includegraphics[width=0.3\textwidth]{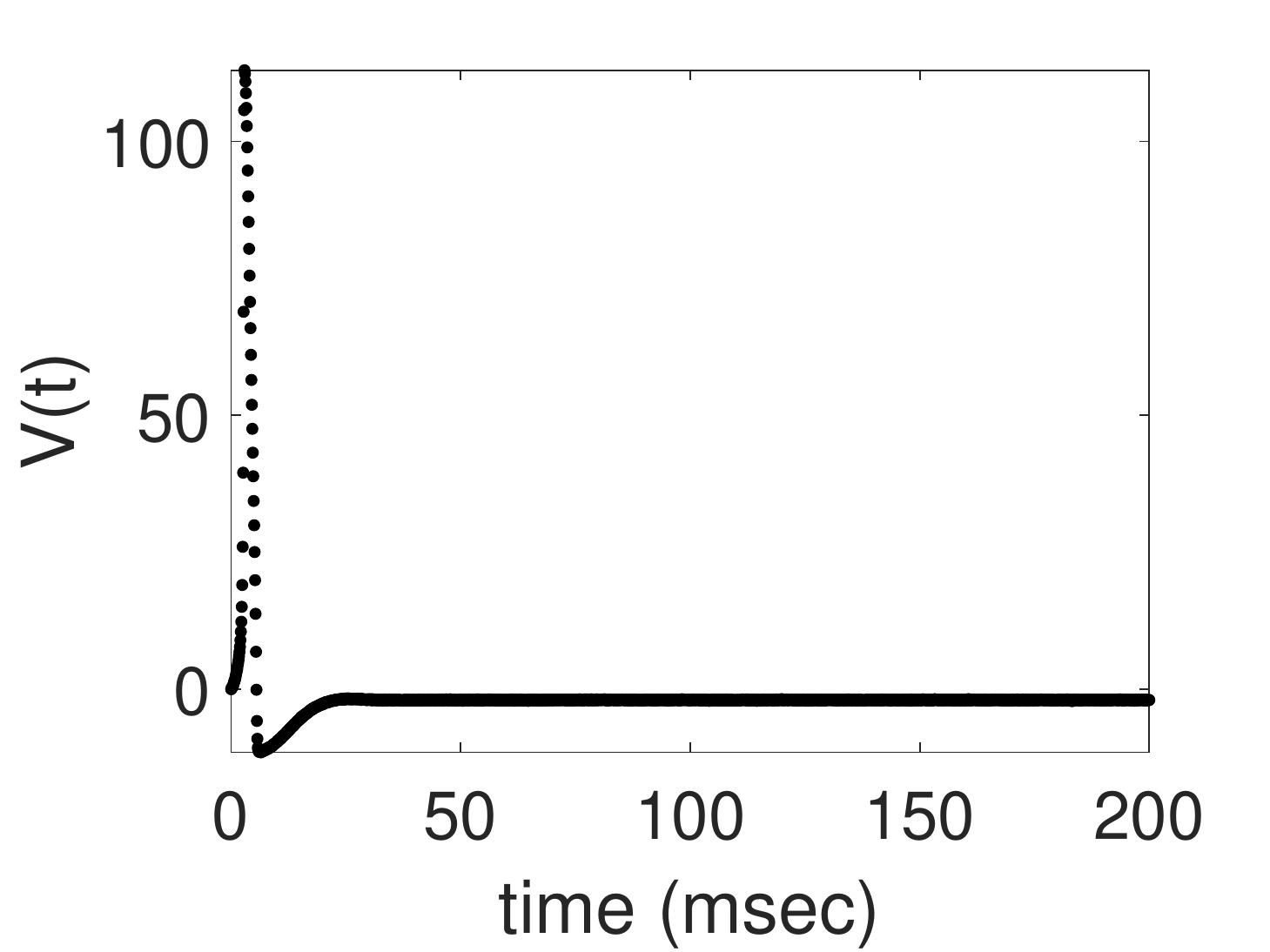}  \includegraphics[width=0.3\textwidth]{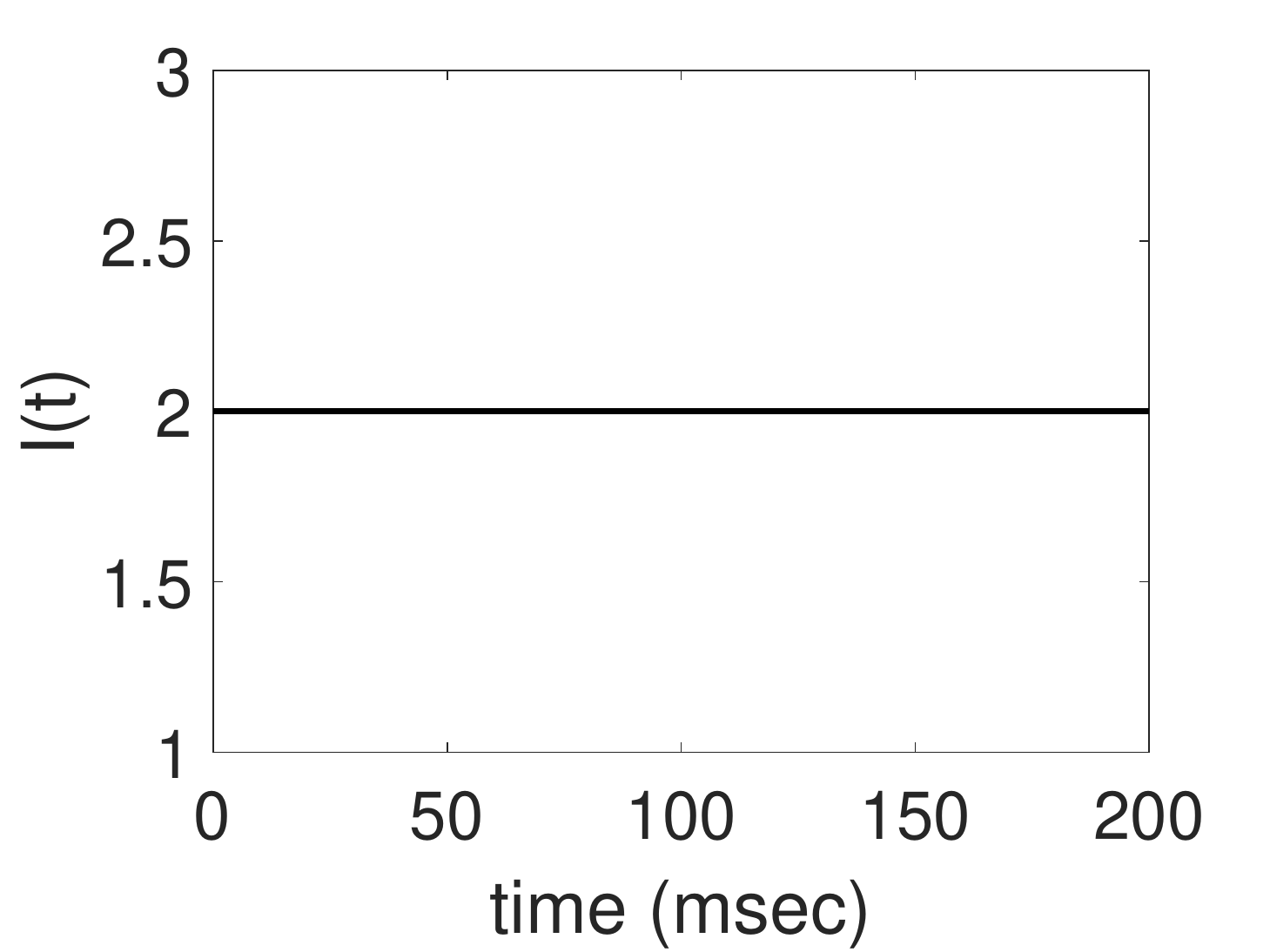}}
\vspace{.2cm}
\centerline{\includegraphics[width=0.3\textwidth]{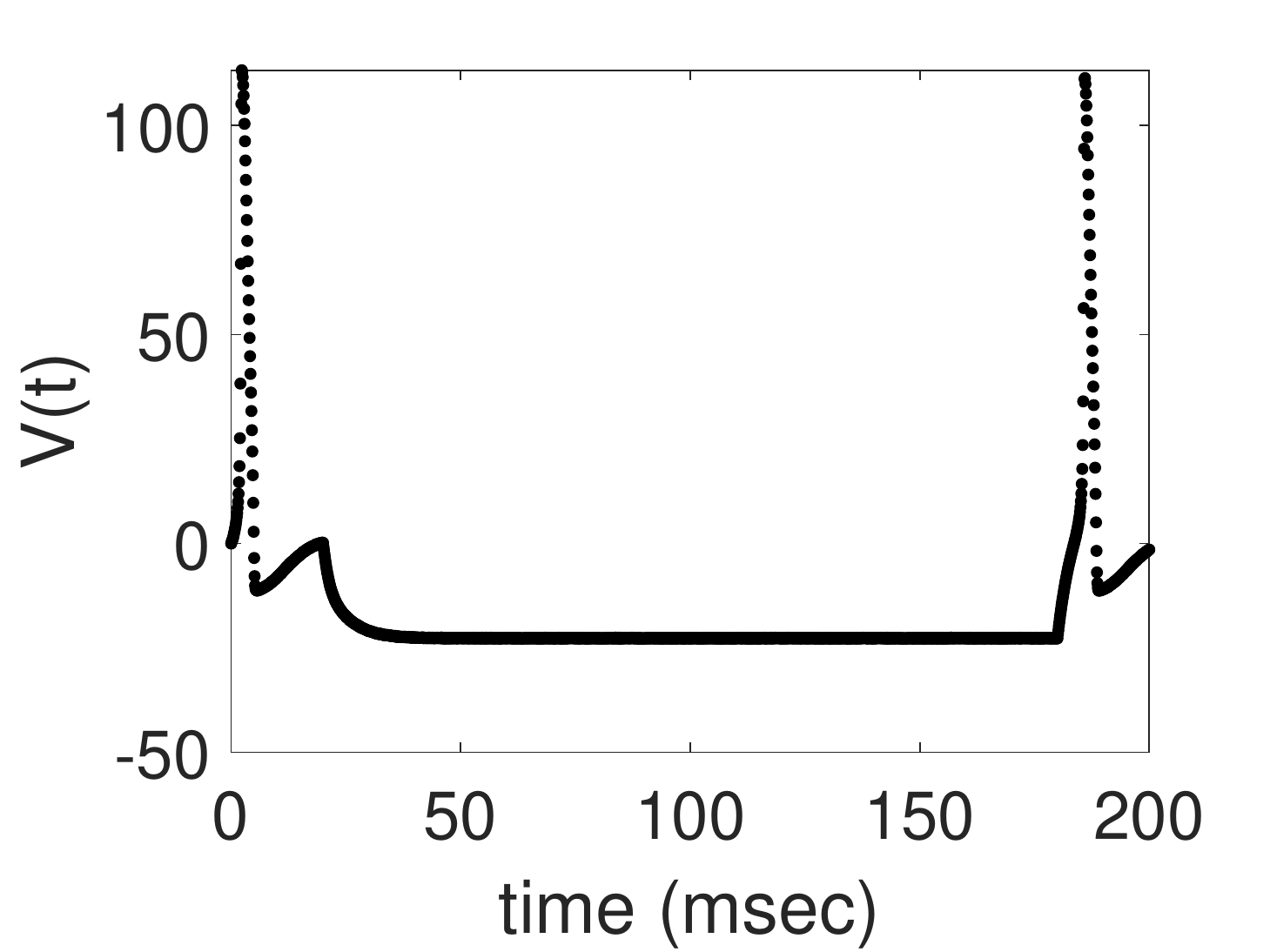}  \includegraphics[width=0.3\textwidth]{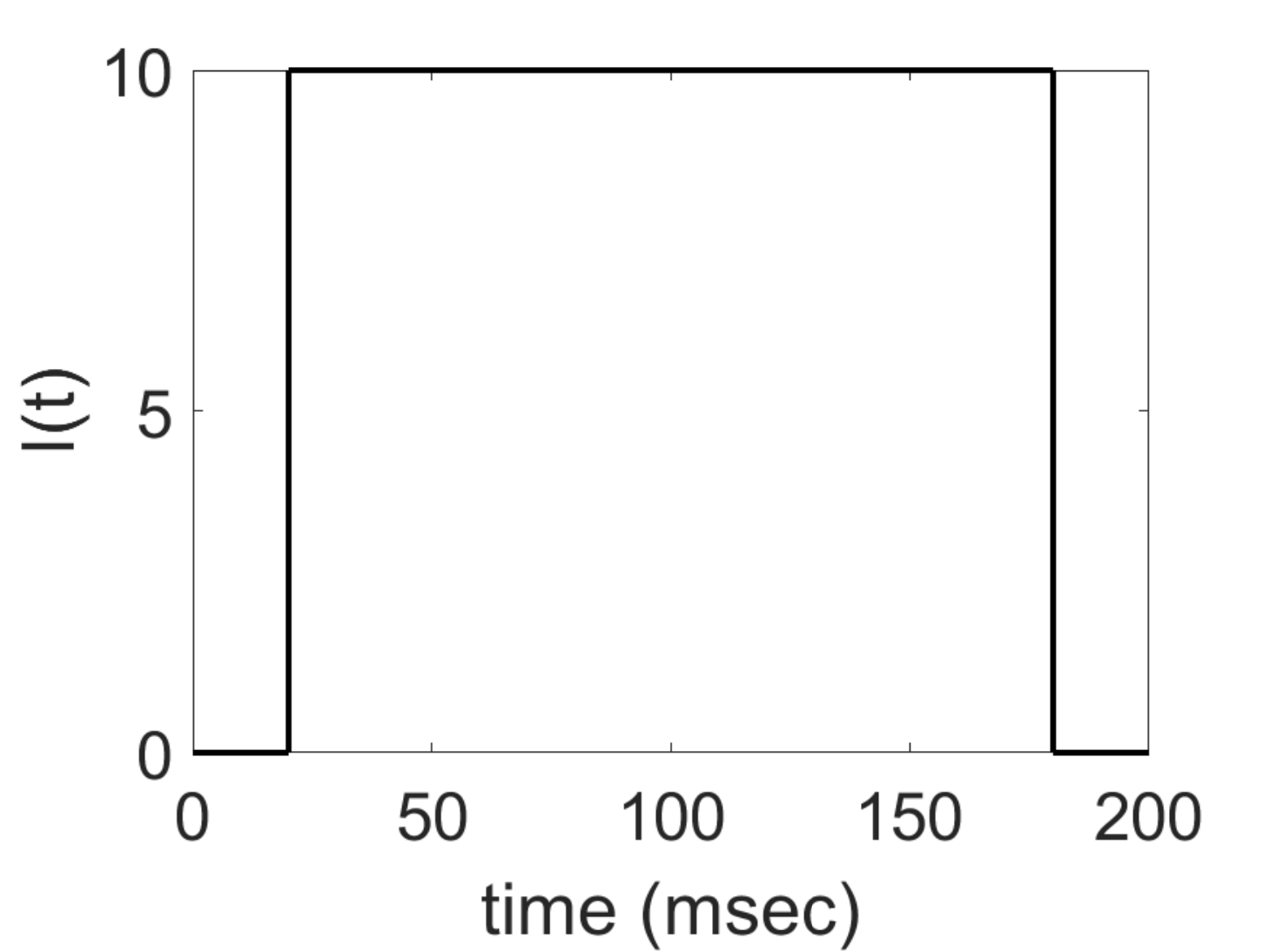}}
\vspace{.2cm}
\centerline{\includegraphics[width=0.3\textwidth]{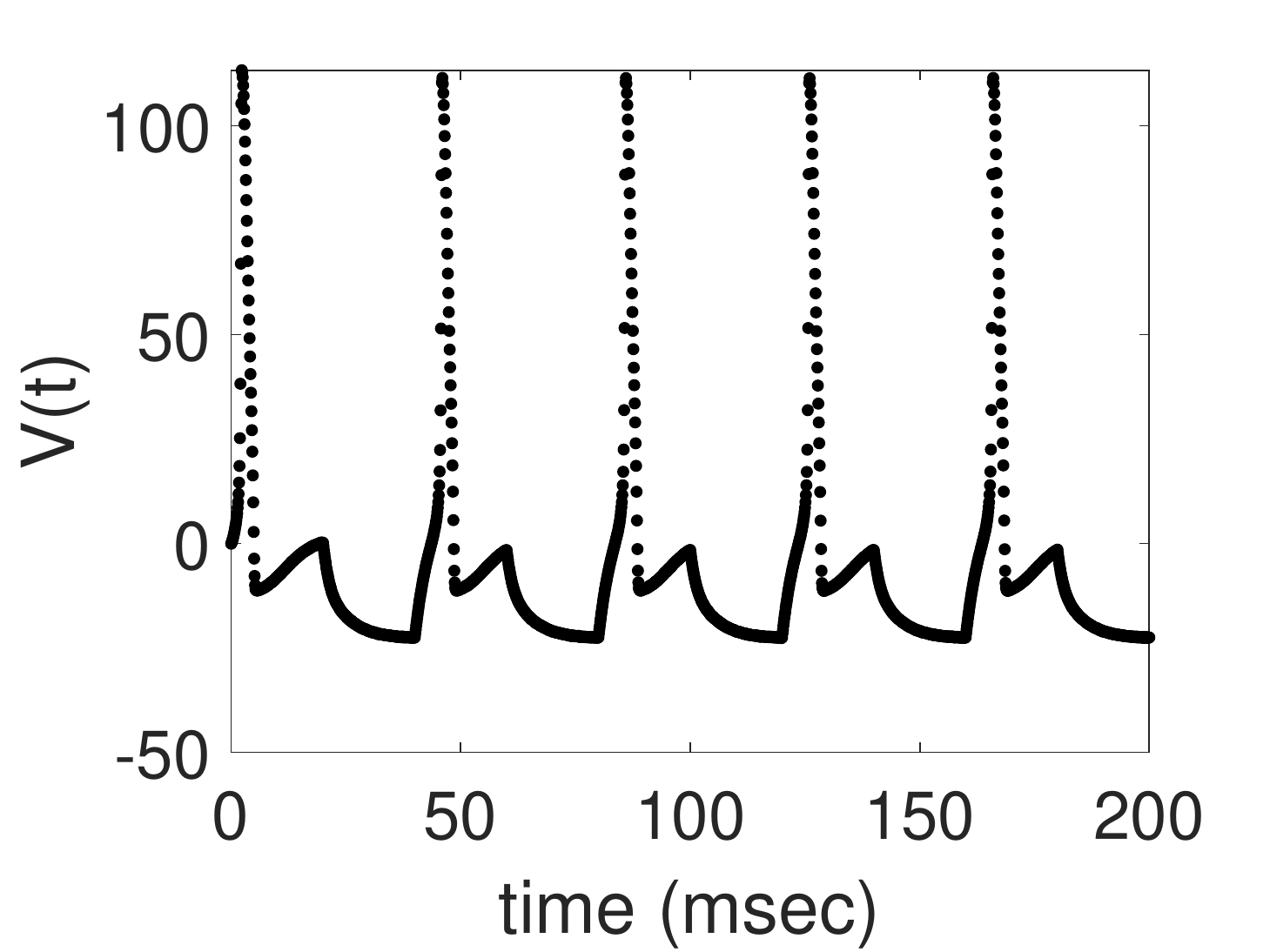}  \includegraphics[width=0.3\textwidth]{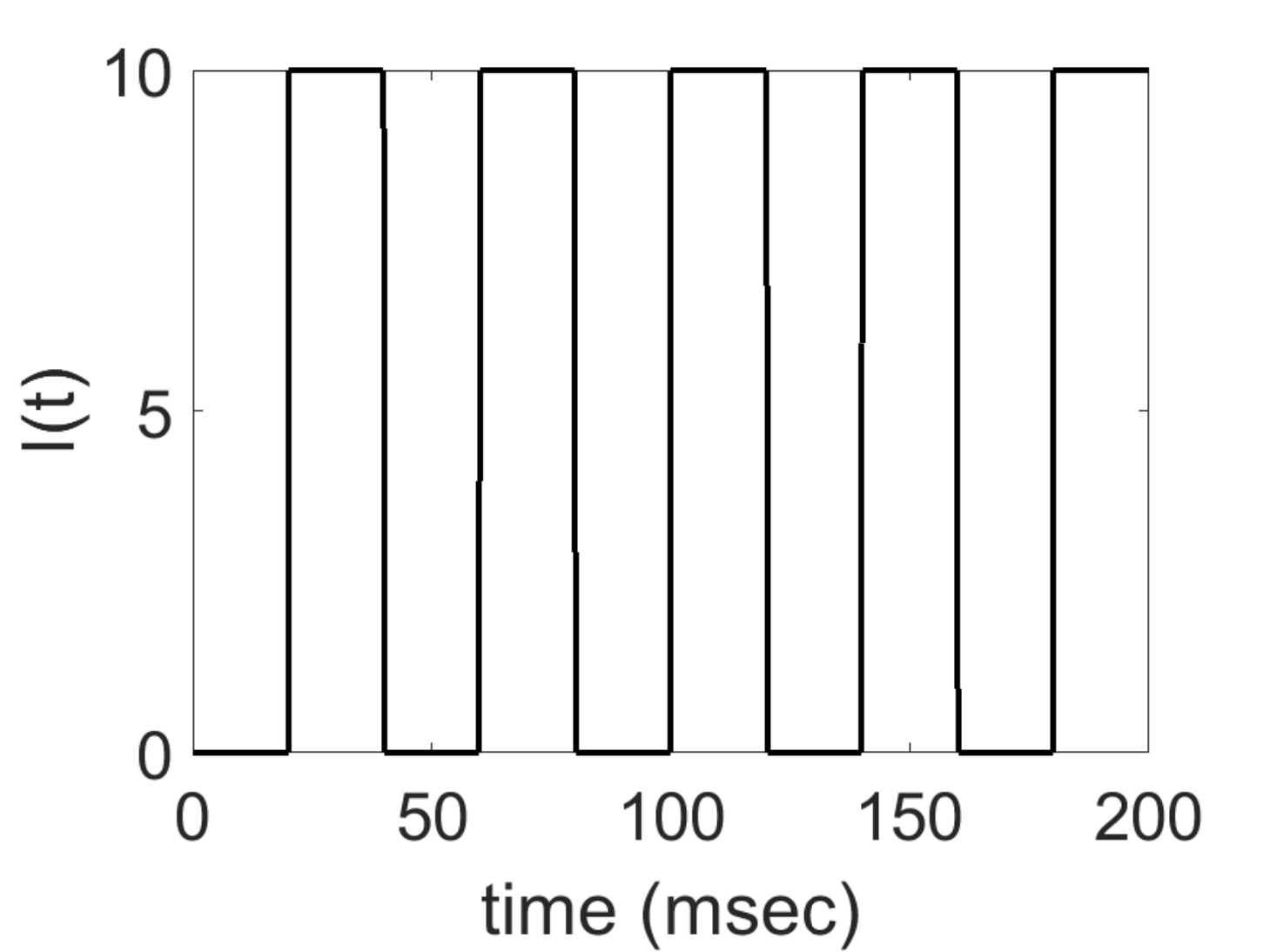}}
\vspace{.2cm}
\centerline{\includegraphics[width=0.3\textwidth]{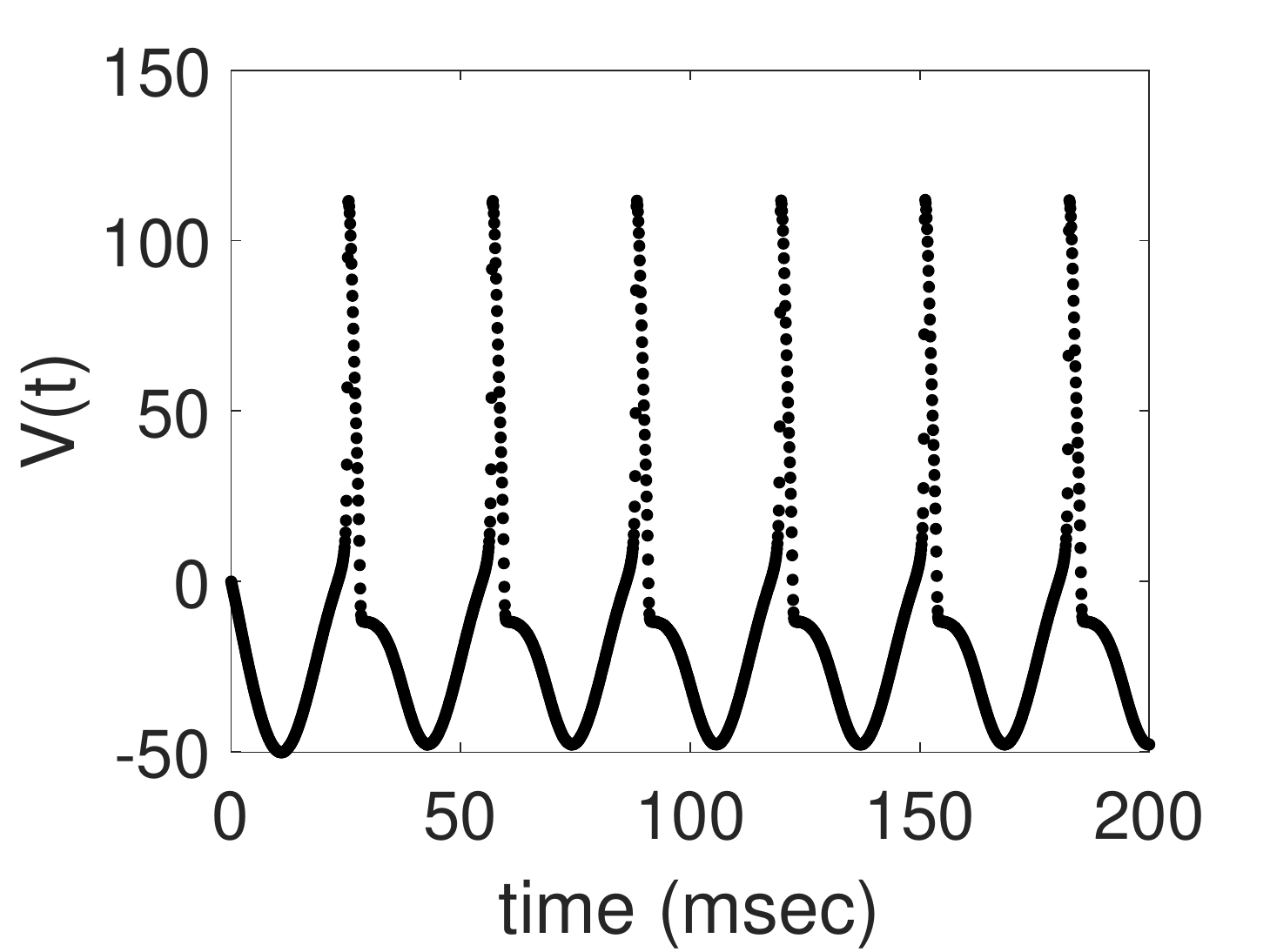}  \includegraphics[width=0.3\textwidth]{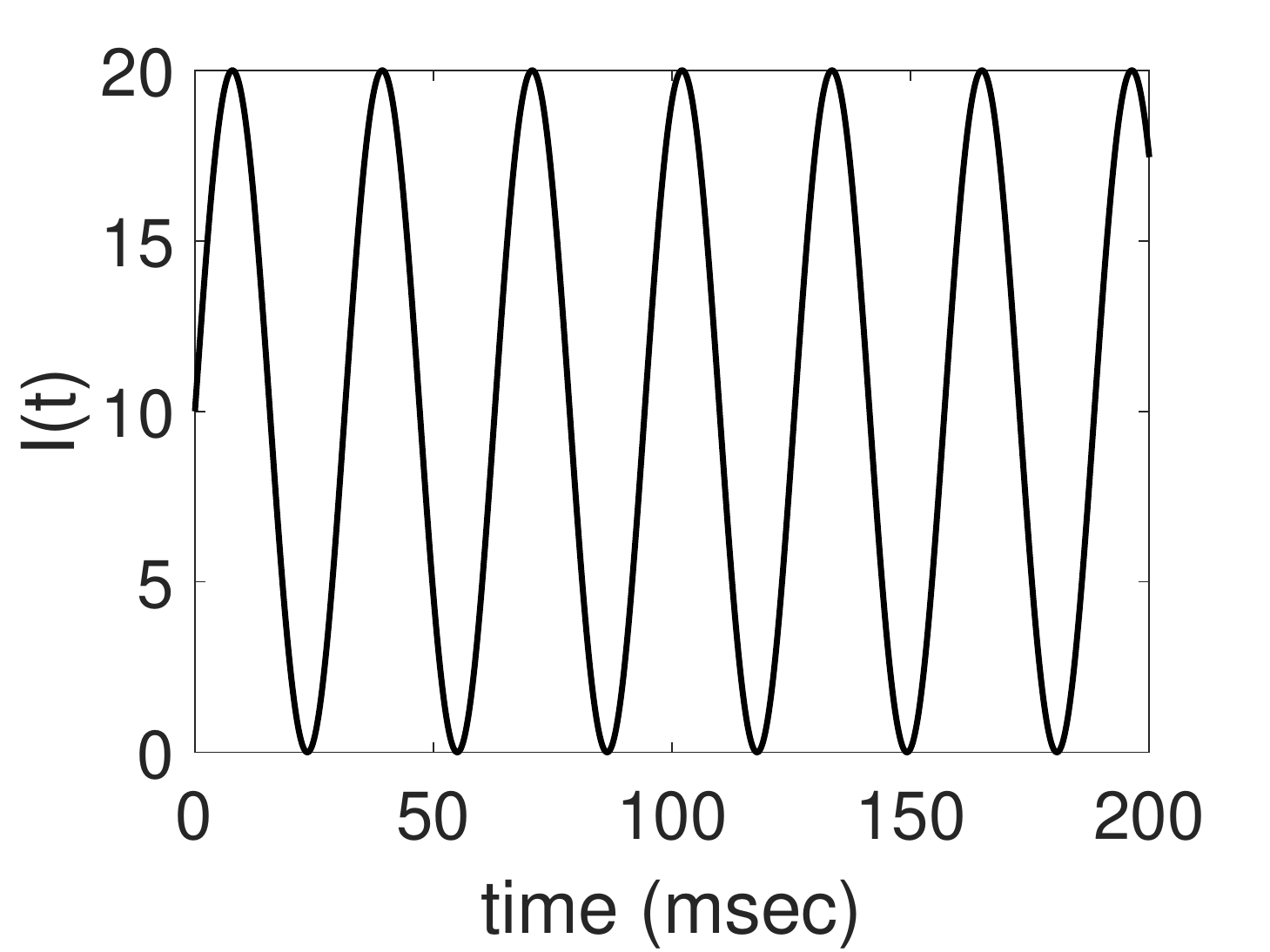}}
\caption{Synthetic measurements of voltage $V(t)$ generated using four different deterministic applied currents $I(t)$.  In each row, the plots on the left depict the noisy voltage measurements generated using the applied currents on the right.  From top to bottom, the plots show data generated using the following current functions: constant current with $I(t)=2$ as in \eqref{Eq:const_curr}; one long step function, defined in \eqref{Eq:longstep_curr}; ``pulsing" step function that alternates every 20 msec, as defined in \eqref{Eq:pulsing_curr}; and the sinusoidal function $I(t) = 10\sin(0.2t) + 10$ as in \eqref{Eq:sine_curr}. }
\label{Fig:Data_and_currents}
\end{figure*}

To generate synthetic data, four different deterministic functions for the applied current $I(t)$ were run through the Hodgkin-Huxley equations \eqref{Eq:HH1}--\eqref{Eq:HH4}. The applied currents considered were:
\begin{itemize}
\item[(a)] a constant current, where
\begin{equation}\label{Eq:const_curr}
I(t) = 2 \text{ mA/cm\textsuperscript{2}} 
\end{equation}
\item[(b)] a step function with one long step, such that 
\begin{equation}\label{Eq:longstep_curr}
I(t) \ = \ \begin{cases} 0  \text{ mA/cm\textsuperscript{2}} ,  &  t \in [0,20) \\[1em]  10  \text{ mA/cm\textsuperscript{2}}  ,  &  t \in [20, 160) \\[1em]  0  \text{ mA/cm\textsuperscript{2}}  ,  &  t \in [160,200]  \end{cases} 
\end{equation}
\item[(c)] a ``pulsing" step function with multiple short steps, such that 
\begin{equation}\label{Eq:pulsing_curr}
I(t) \ = \ \begin{cases} 0 \text{ mA/cm\textsuperscript{2}}  , & t \in [20q, 20 + 20q), \\ 
\ & \quad q = 0, 2, 4, 6, 8\\[1em]
10  \text{ mA/cm\textsuperscript{2}} , & t \in [20q, 20 + 20q), \\
\ & \quad q = 1, 3, 5, 7, 9 \end{cases} 
\end{equation}
\item[(d)] a sinusoidal function, where
\begin{equation}\label{Eq:sine_curr}
 I(t) = 10\sin(0.2t) + 10 .
\end{equation} 
\end{itemize}
For each data set, measurements of voltage $V(t)$ were taken at 2,001 equidistant time instances over the interval $[0,200]$ and corrupted by Gaussian noise with zero mean and standard deviation 0.05.  Figure \ref{Fig:Data_and_currents} shows each data sets along with the corresponding applied current.  Note that the gating variables $n(t)$, $m(t)$, and $h(t)$ are unobserved states.

\subsection{Estimating Time-Varying Applied Current via Parameter Tracking}

\begin{figure*}[t!]
\centerline{\includegraphics[width=0.25\textwidth]{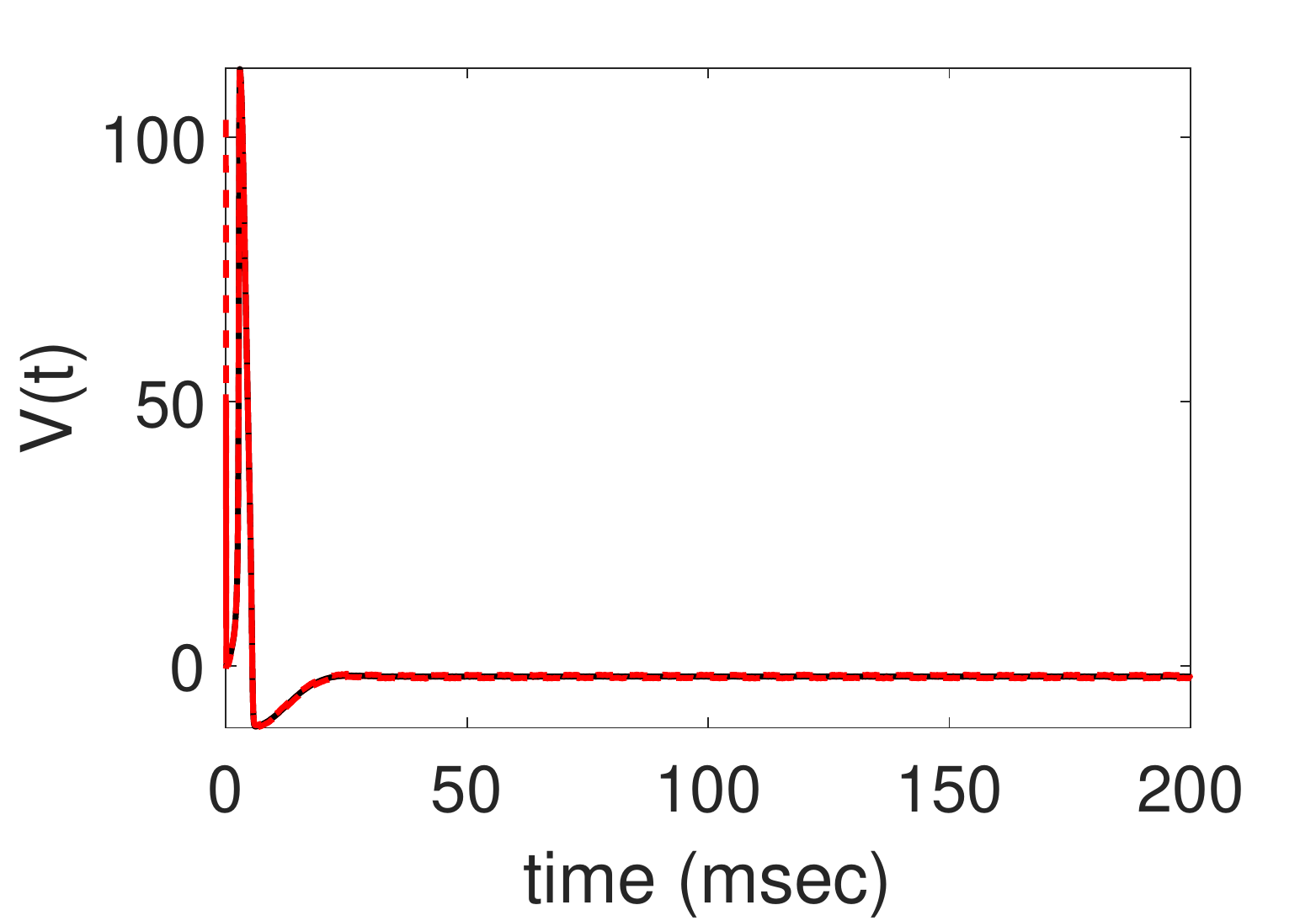} \includegraphics[width=0.25\textwidth]{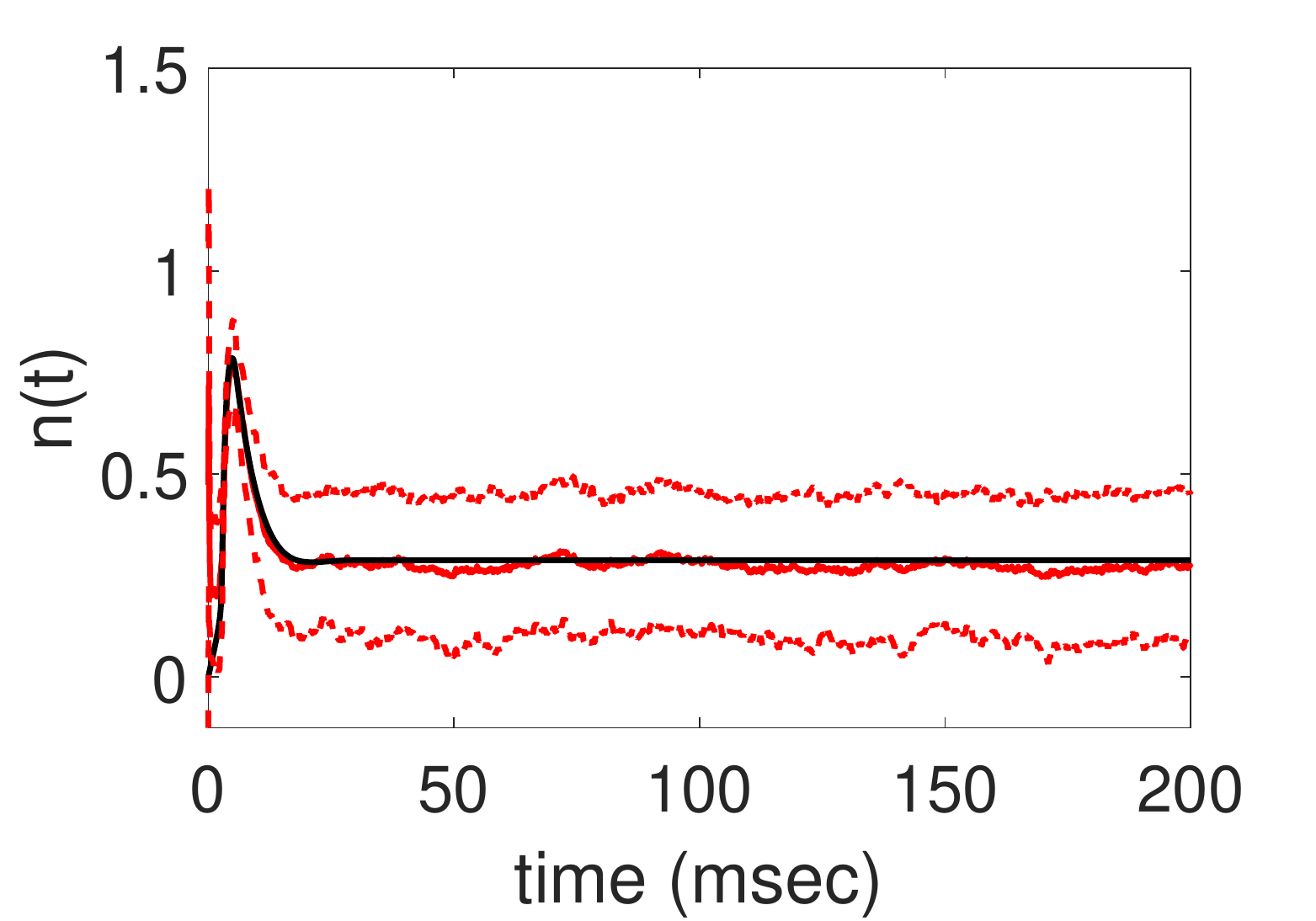}  \includegraphics[width=0.25\textwidth]{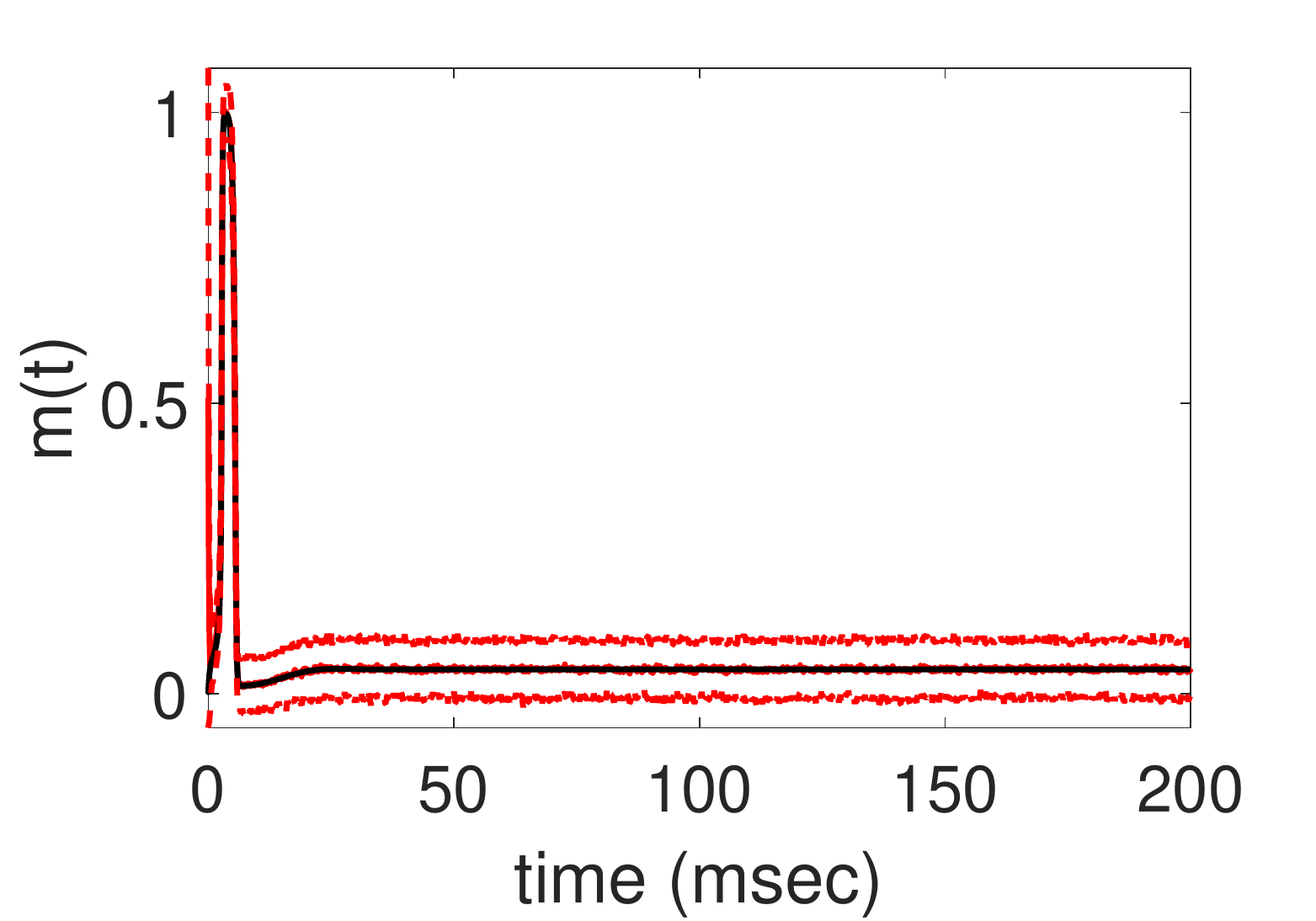} \includegraphics[width=0.25\textwidth]{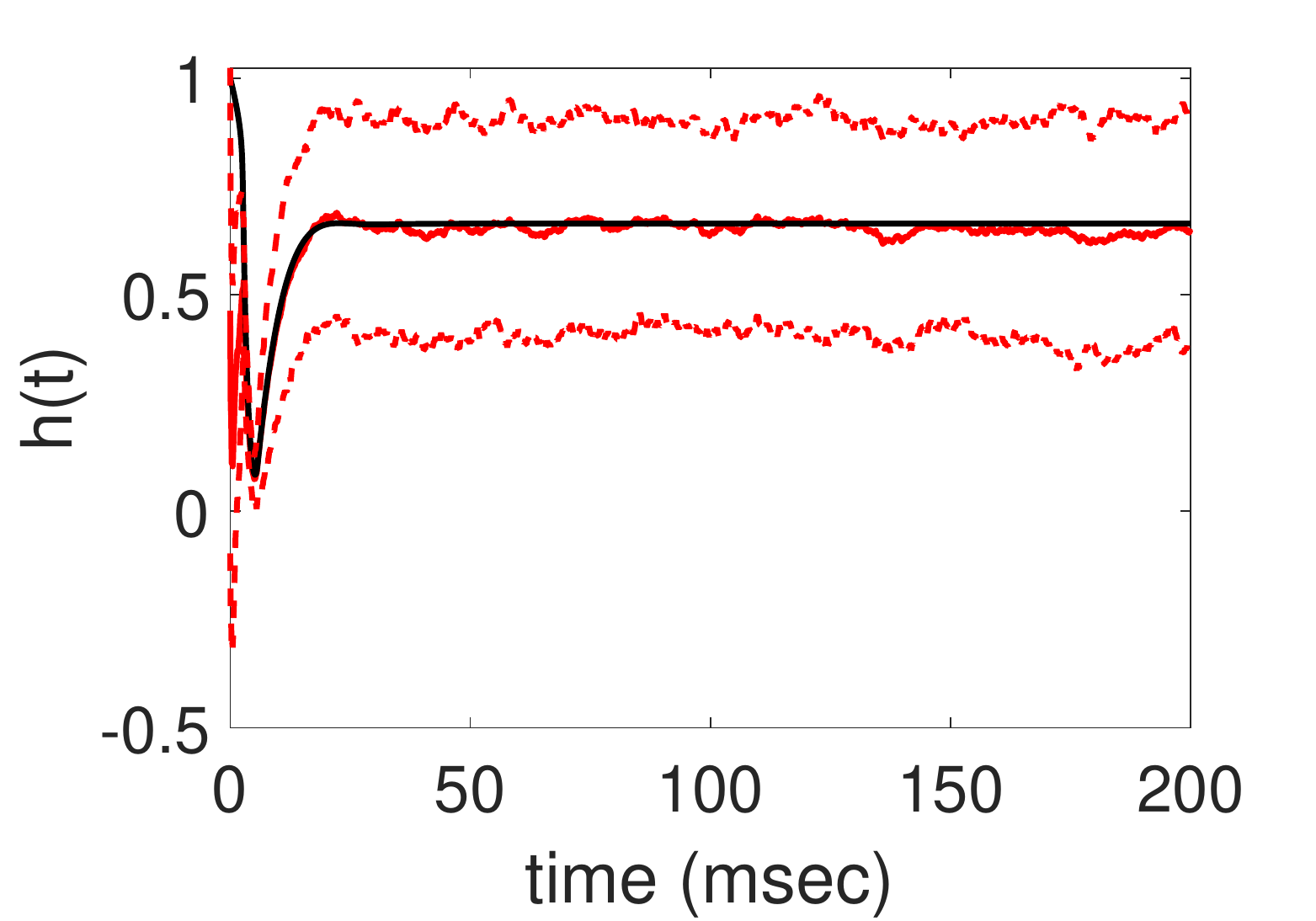} }
\vspace{.4cm}
\centerline{\includegraphics[width=0.75\textwidth]{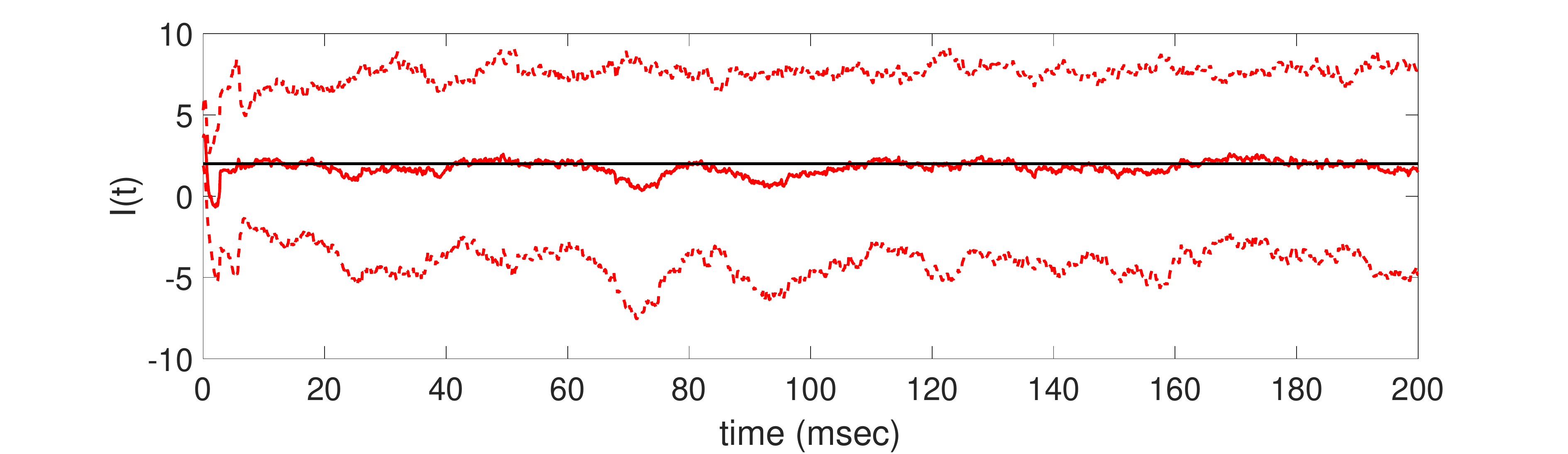} }
\caption{Resulting EnKF with parameter tracking estimates of $V(t)$, $n(t)$, $m(t)$, $h(t)$ (top, from left to right), and applied current $I(t)$ (bottom) from the data obtained from the constant current $I(t) = 2$ mA/cm\textsuperscript{2} in \eqref{Eq:const_curr}. In each panel, the EnKF estimated mean is shown in solid red while the true solution is shown in solid black. The dashed red lines show the estimated $\pm 2$ standard deviation curves around the mean.}
\label{Fig:constant_final}
\end{figure*}

\begin{figure*}[t!]
\centerline{\includegraphics[width=0.25\textwidth]{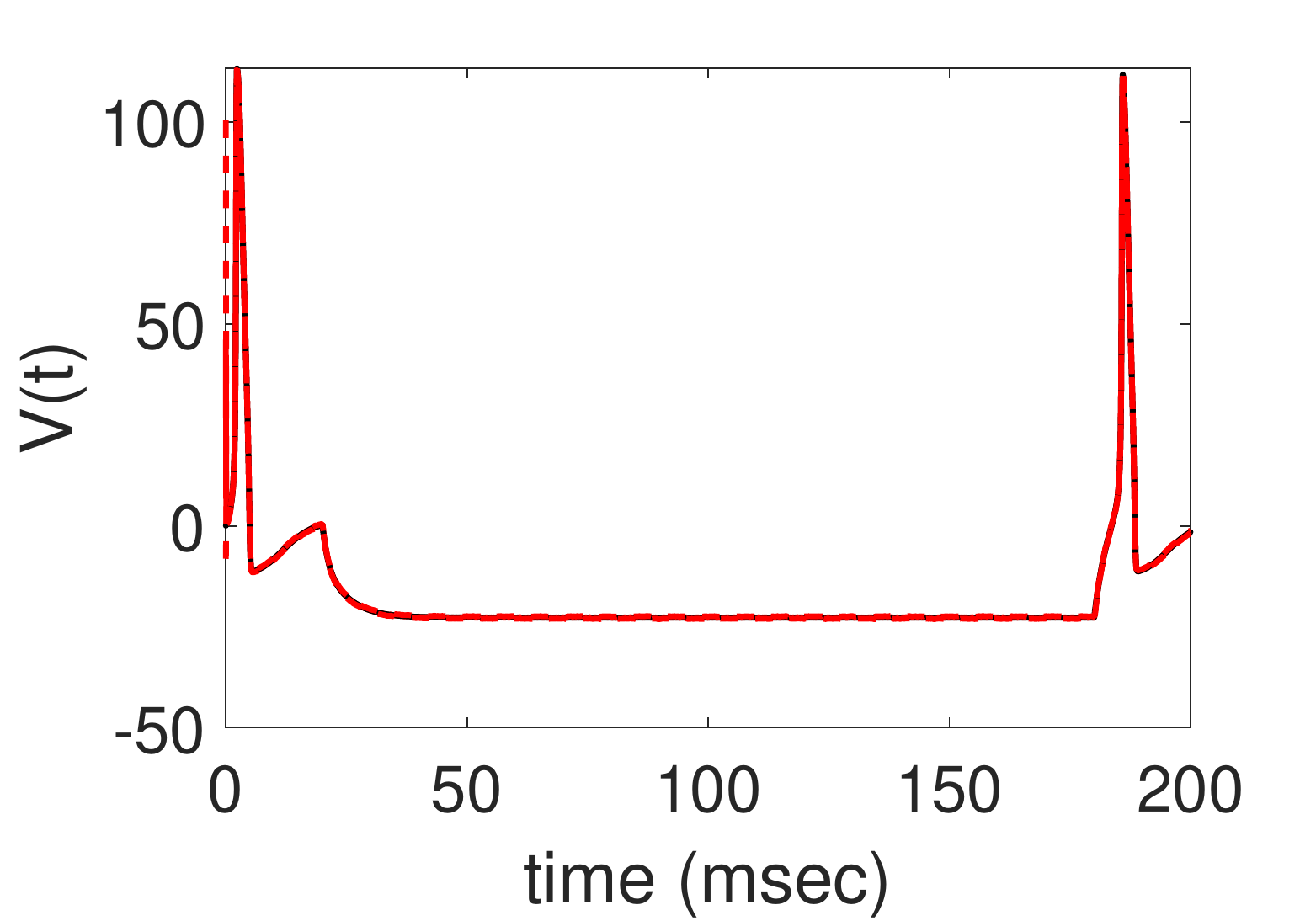} \includegraphics[width=0.25\textwidth]{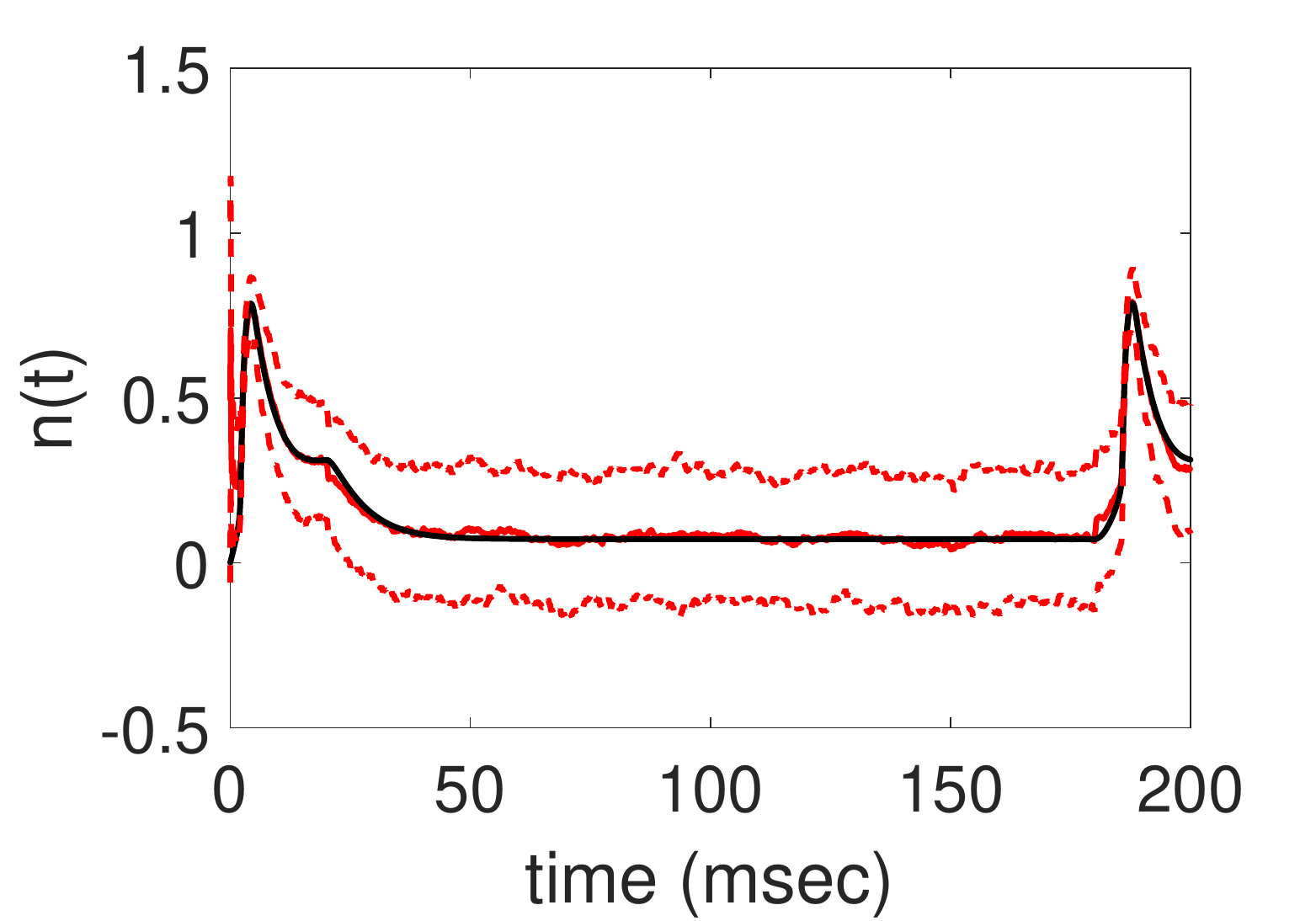} 
\includegraphics[width=0.25\textwidth]{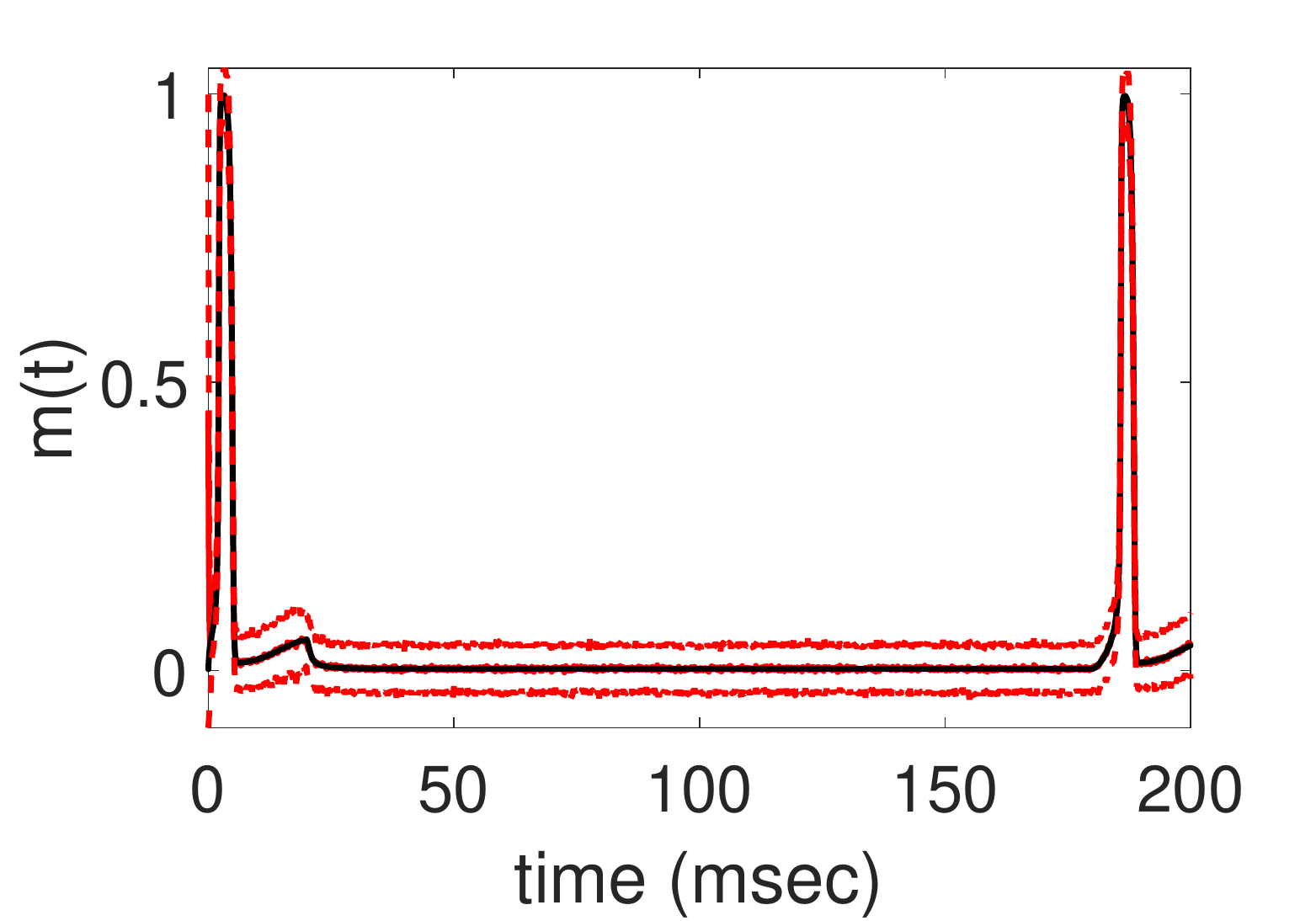} \includegraphics[width=0.25\textwidth]{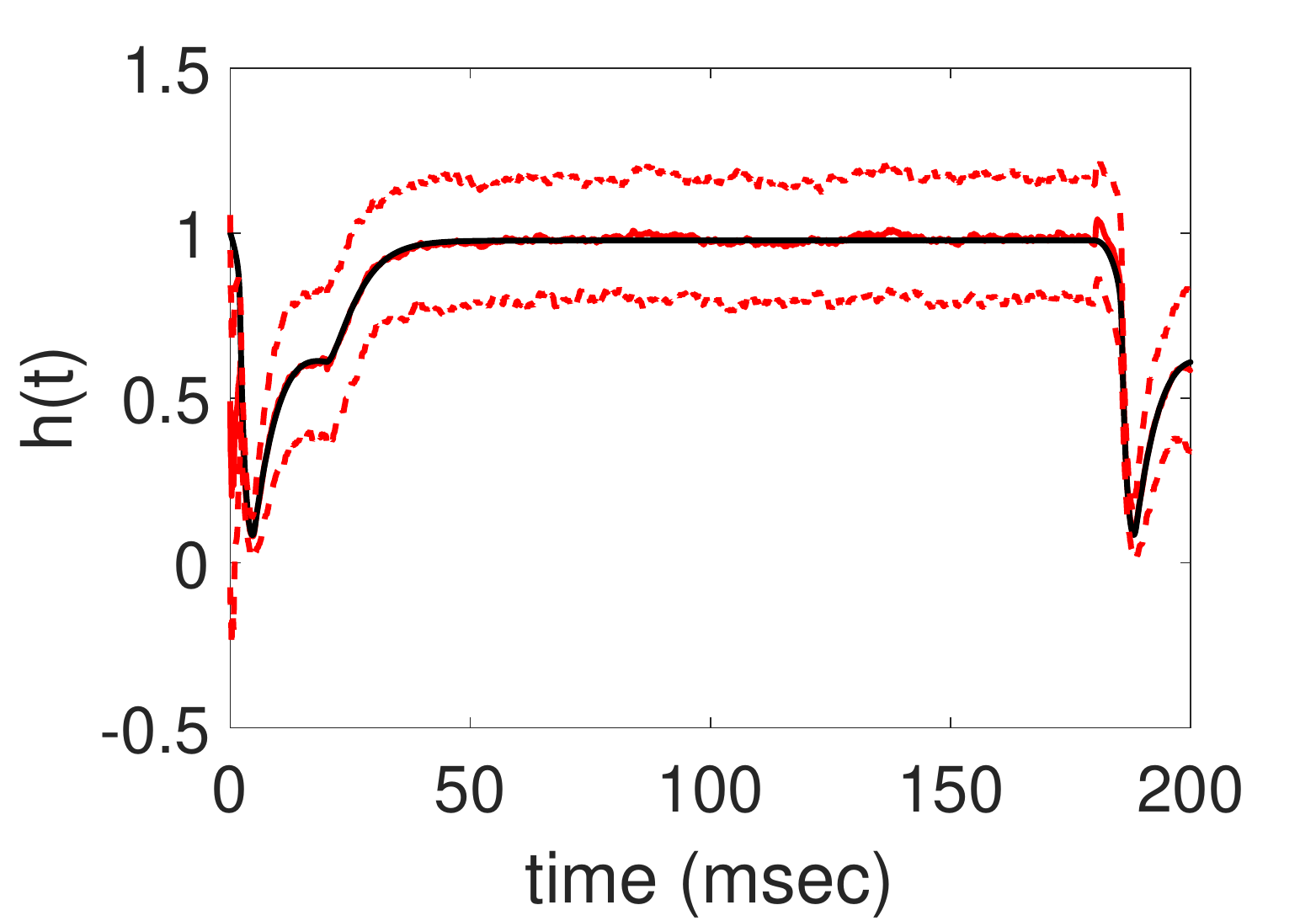} }
\vspace{.4cm}
\centerline{\includegraphics[width=0.75\textwidth]{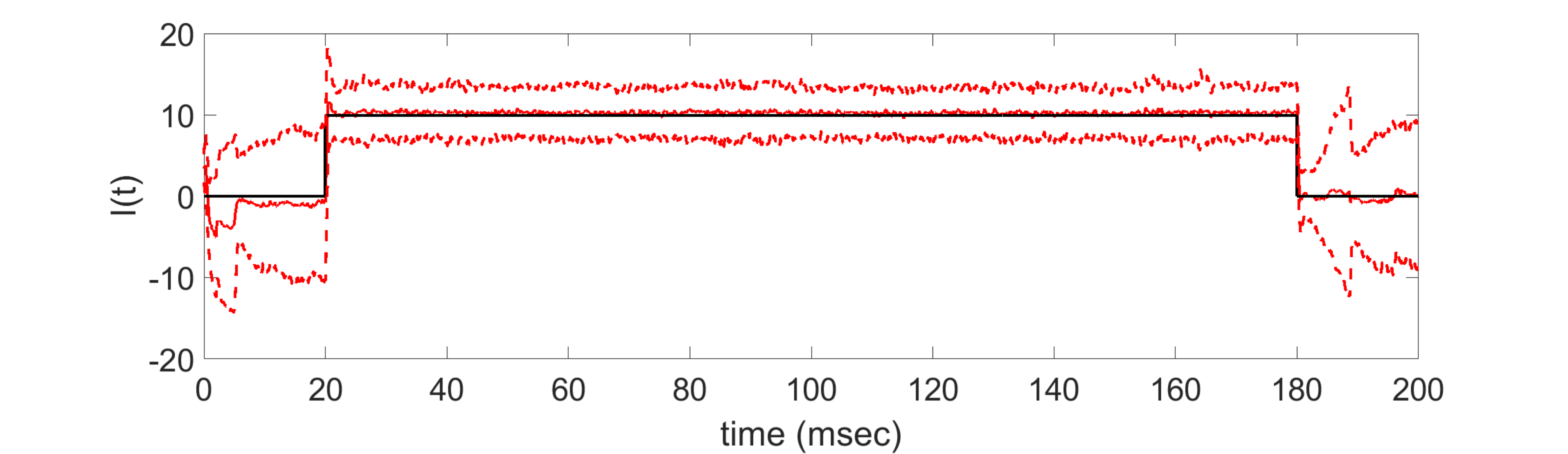} }
\caption{Resulting EnKF with parameter tracking estimates of $V(t)$, $n(t)$, $m(t)$, $h(t)$ (top, from left to right), and applied current $I(t)$ (bottom) from the data obtained from the one-step current in \eqref{Eq:longstep_curr}. In each panel, the EnKF estimated mean is shown in solid red while the true solution is shown in solid black. The dashed red lines show the estimated $\pm 2$ standard deviation curves around the mean.}
\label{Fig:1_long_step}
\end{figure*}

\begin{figure*}[t!]
\centerline{\includegraphics[width=0.25\textwidth]{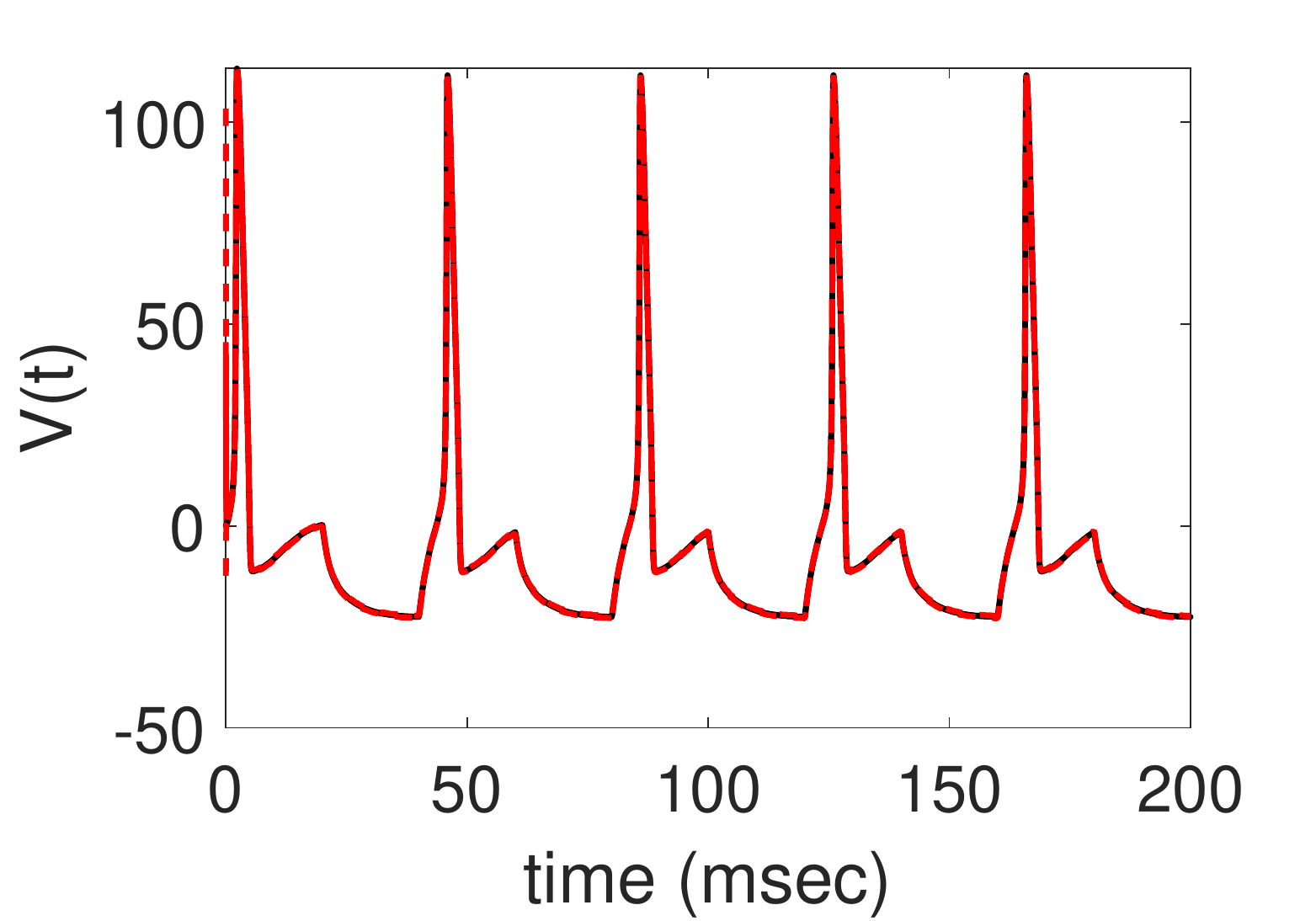} \includegraphics[width=0.25\textwidth]{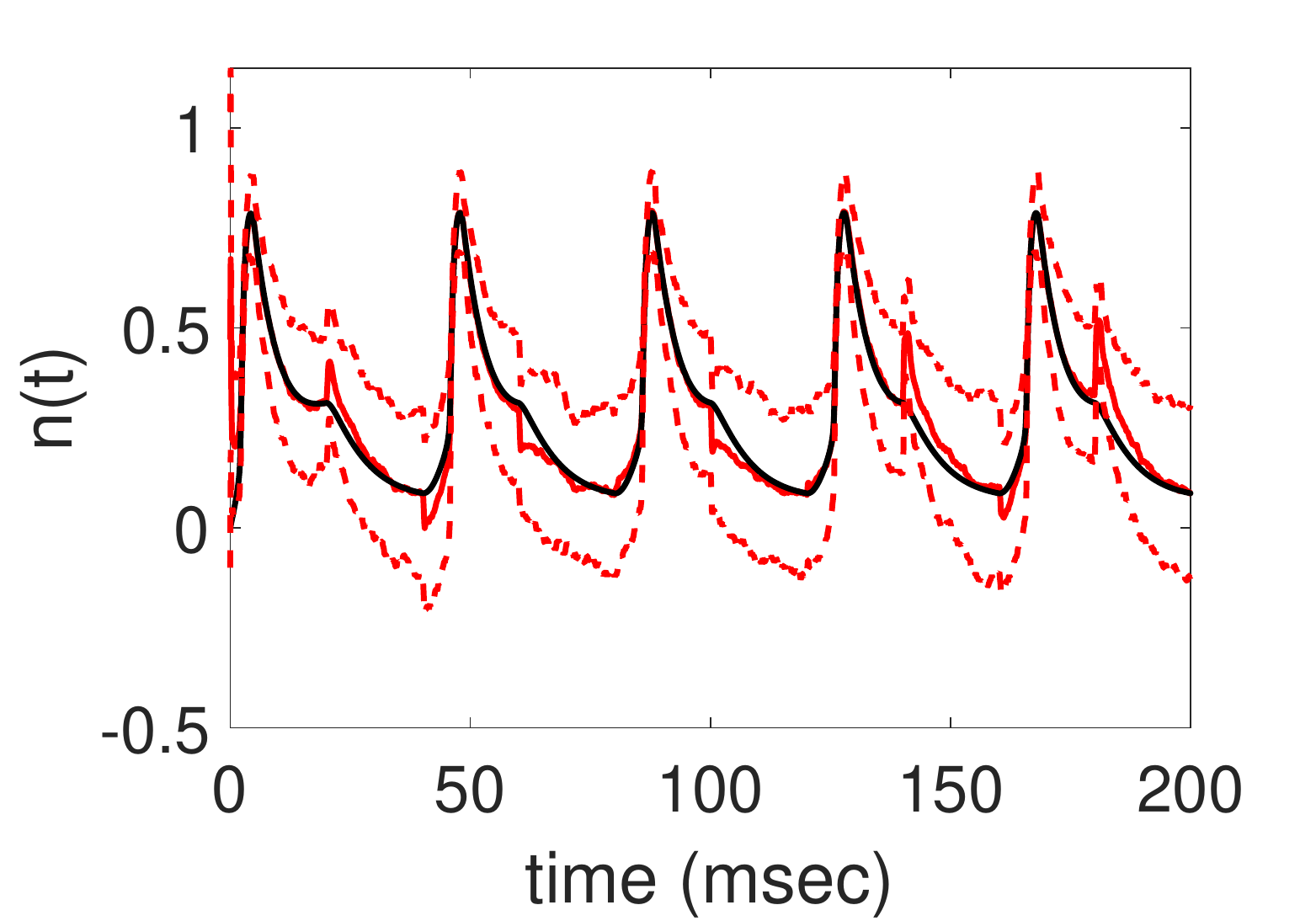} 
\includegraphics[width=0.25\textwidth]{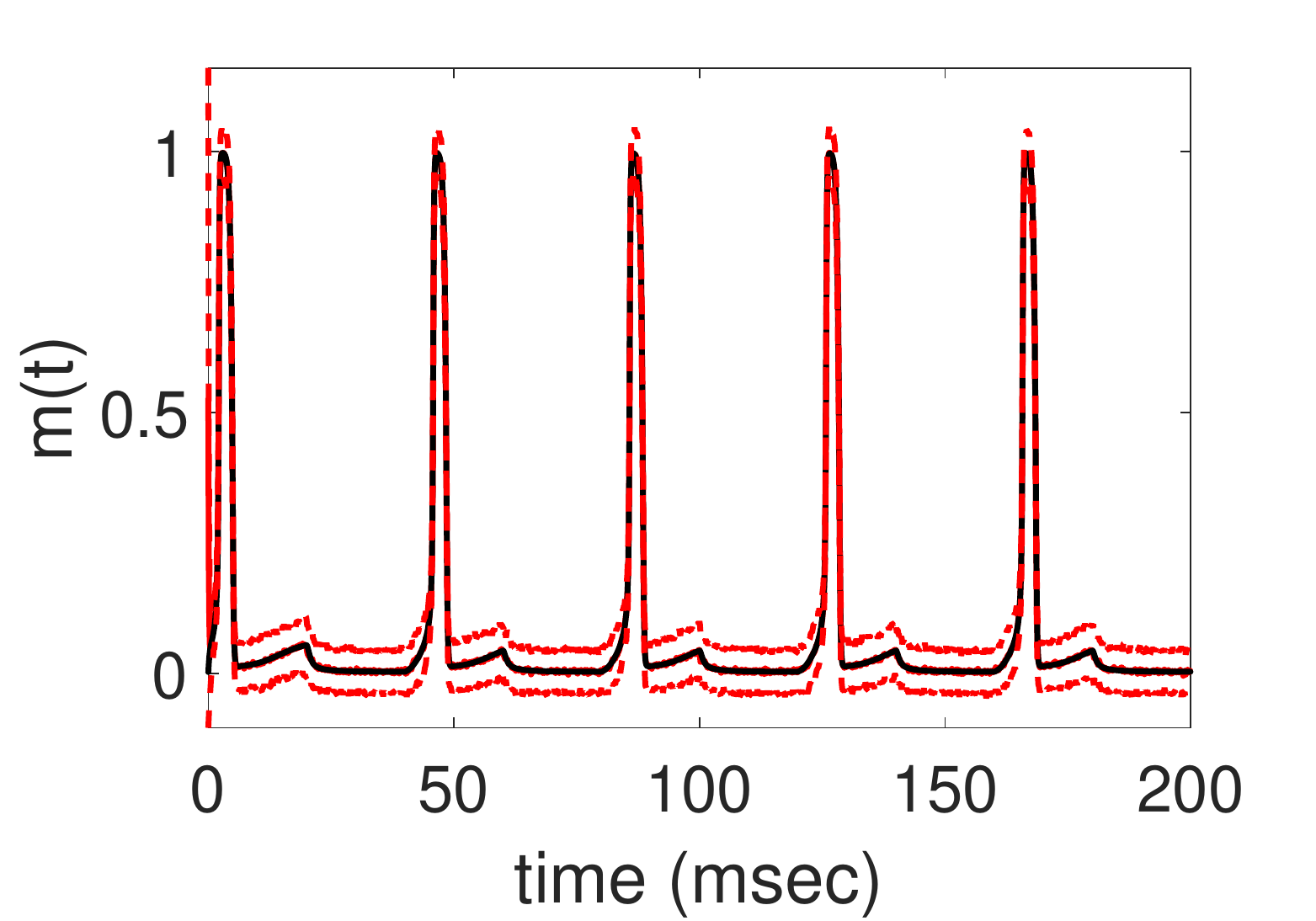} \includegraphics[width=0.25\textwidth]{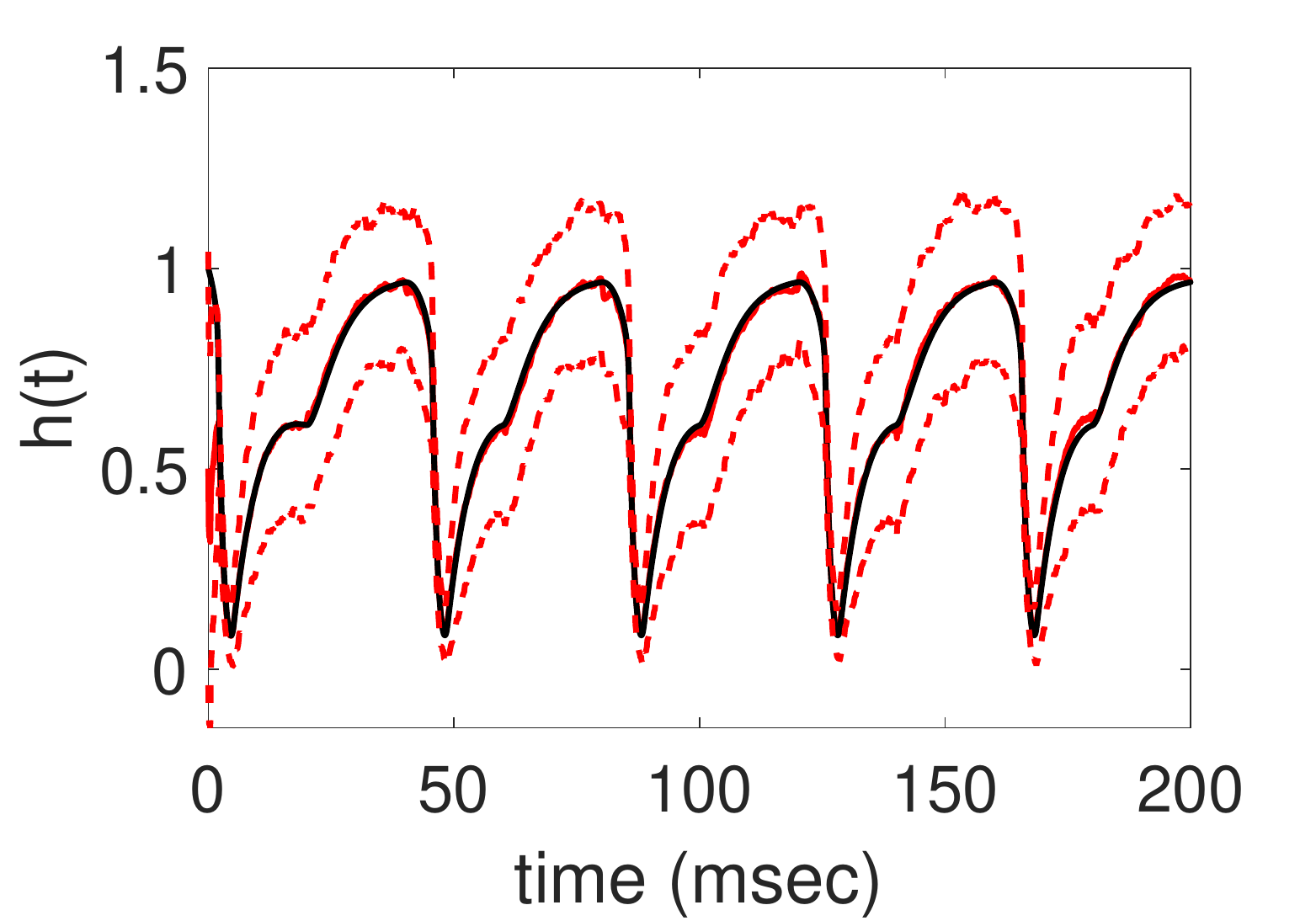} }
\vspace{.4cm}
\centerline{\includegraphics[width=0.75\textwidth]{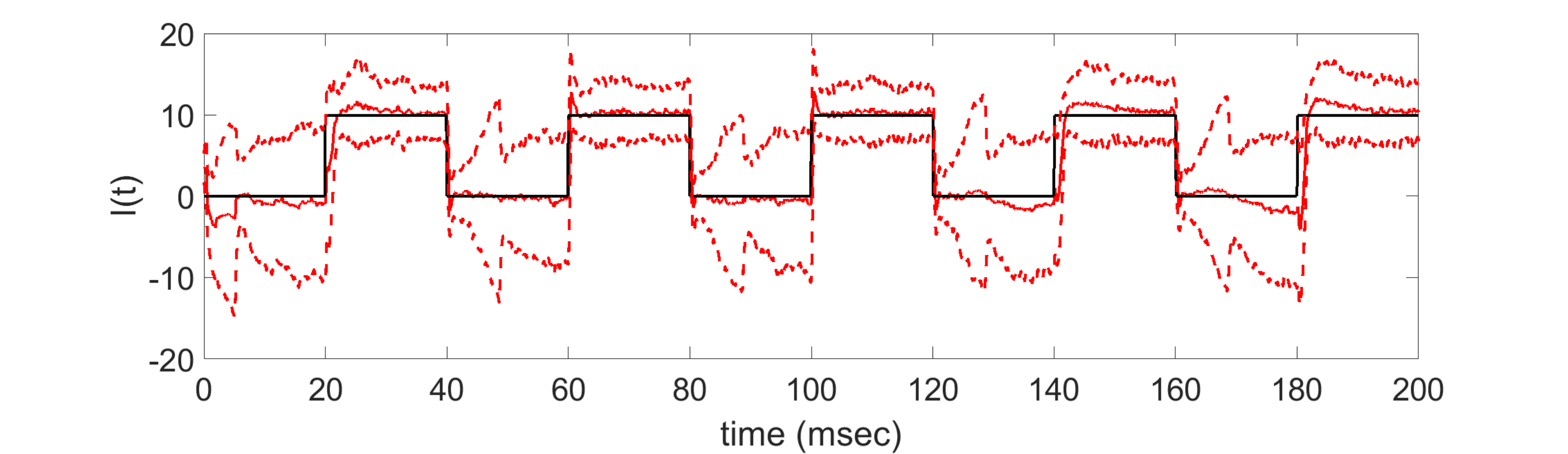} }
\caption{Resulting EnKF with parameter tracking estimates of $V(t)$, $n(t)$, $m(t)$, $h(t)$ (top, from left to right), and applied current $I(t)$ (bottom) from the data obtained from the ``pulsing'' step current in \eqref{Eq:pulsing_curr}. In each panel, the EnKF estimated mean is shown in solid red while the true solution is shown in solid black. The dashed red lines show the estimated $\pm 2$ standard deviation curves around the mean.}
\label{Fig:step_final}
\end{figure*}

\begin{figure*}[t!]
\centerline{\includegraphics[width=0.25\textwidth]{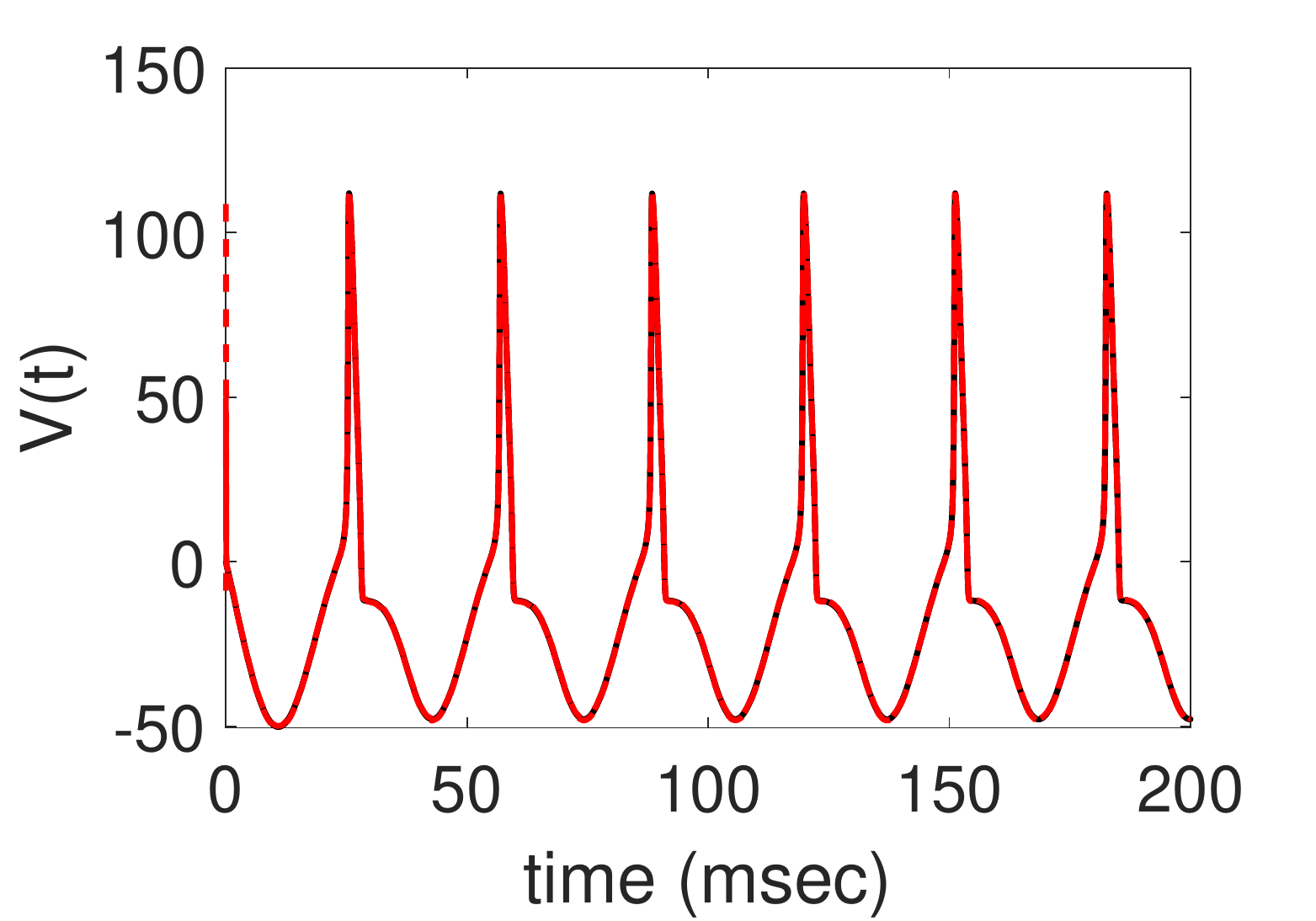} \includegraphics[width=0.25\textwidth]{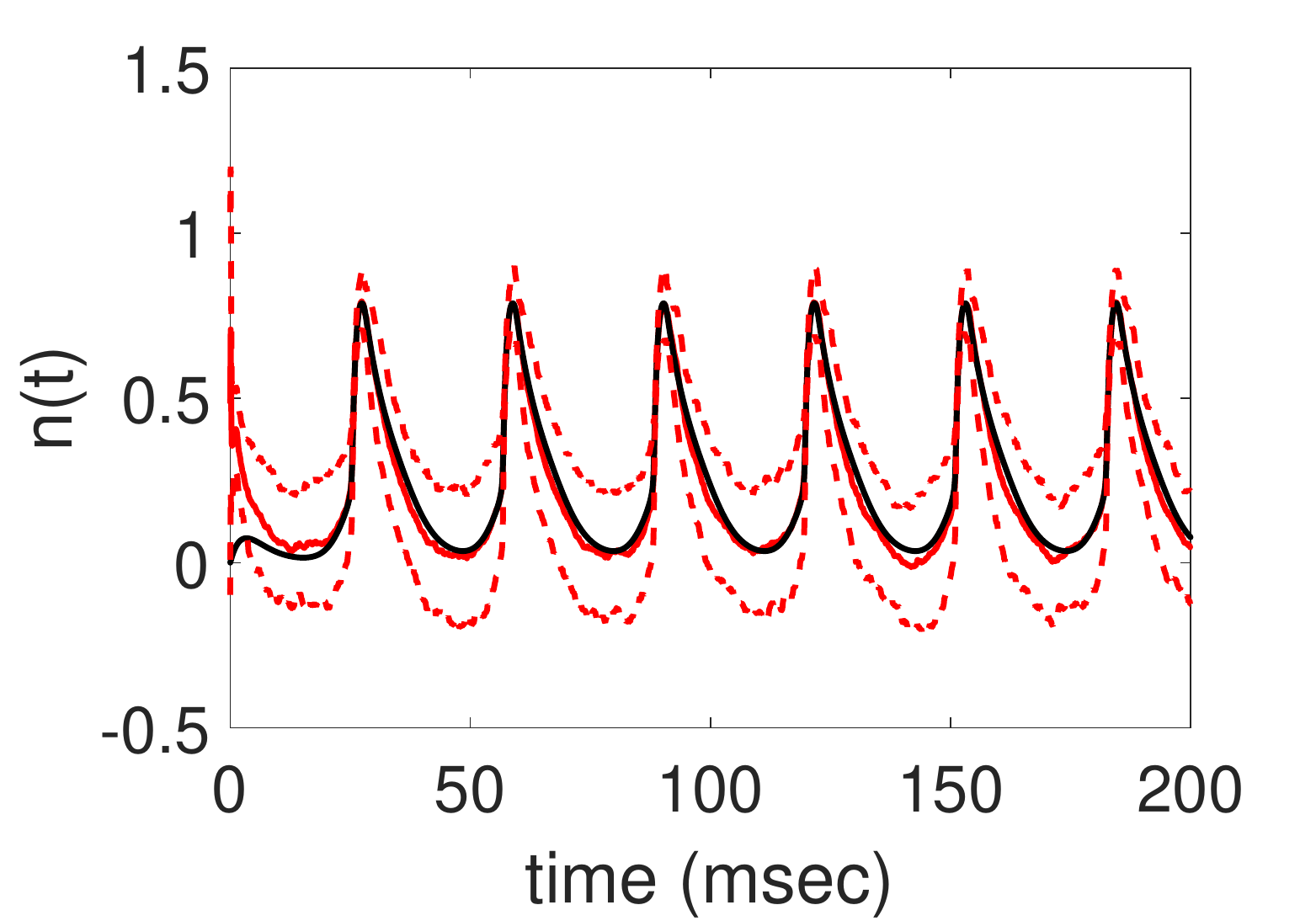} 
\includegraphics[width=0.25\textwidth]{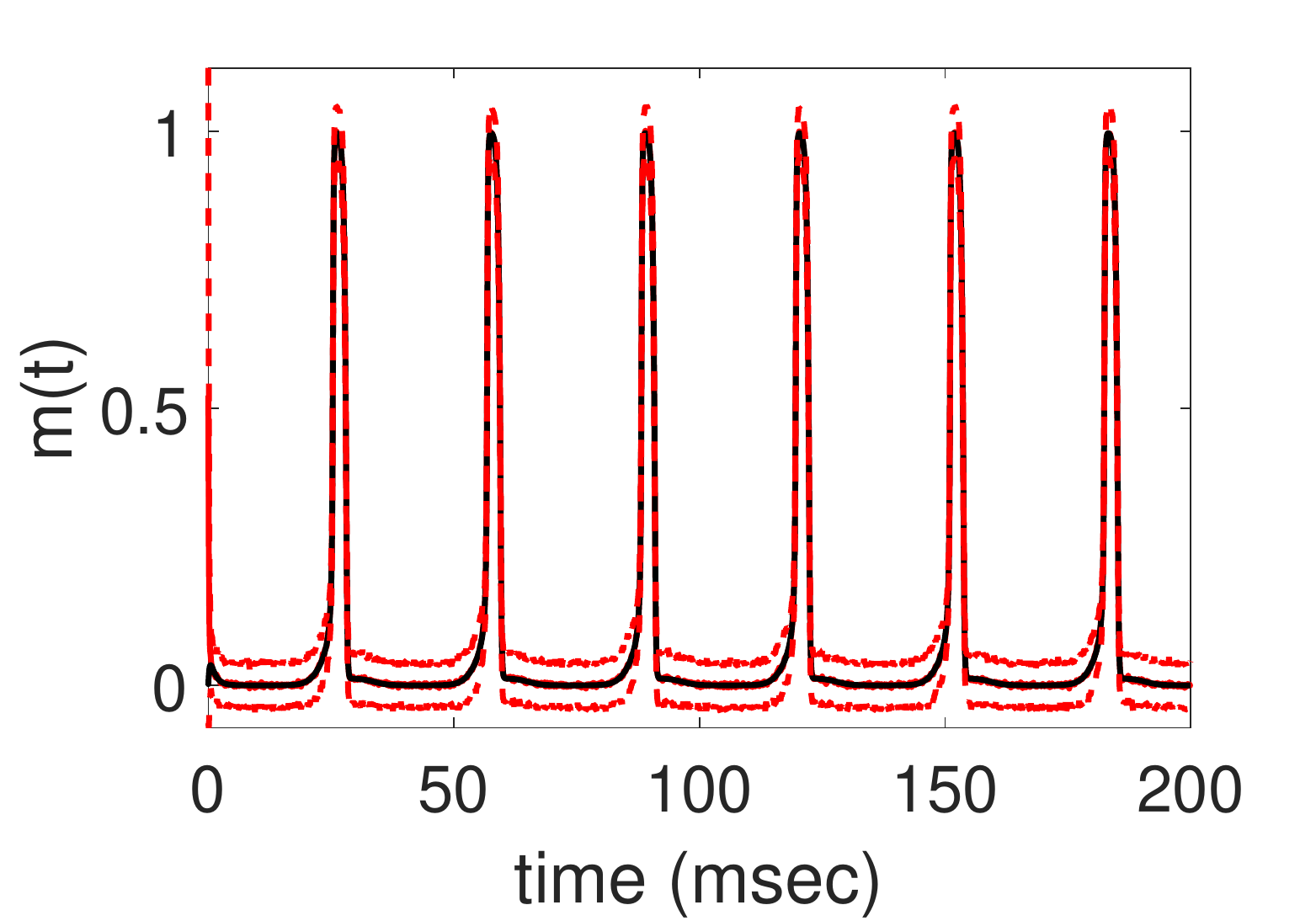} \includegraphics[width=0.25\textwidth]{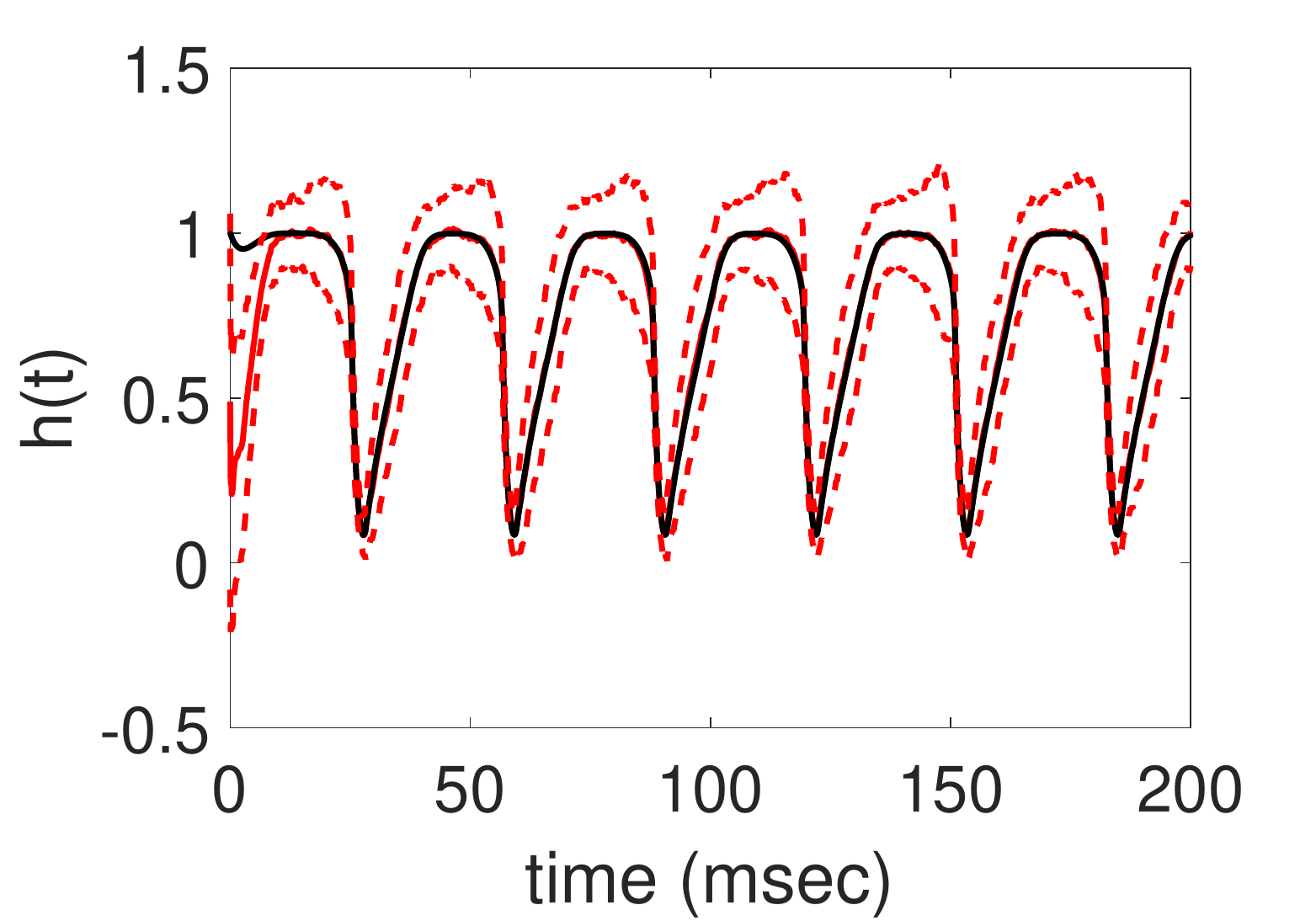} }
\vspace{.4cm}
\centerline{\includegraphics[width=0.75\textwidth]{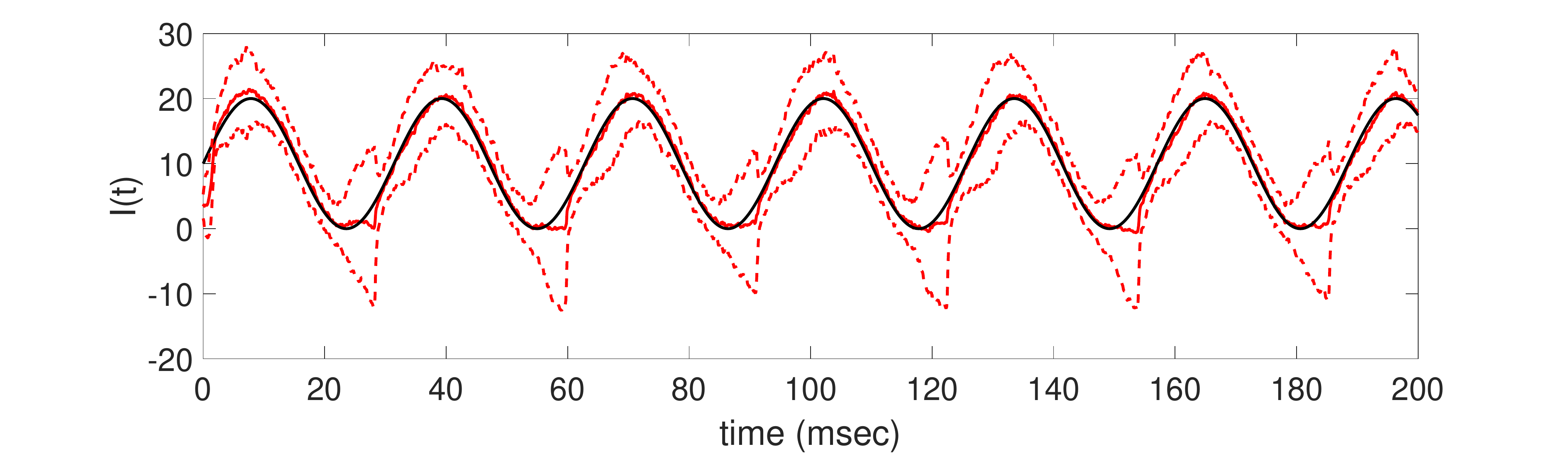} }
\caption{Resulting EnKF with parameter tracking estimates of $V(t)$, $n(t)$, $m(t)$, $h(t)$ (top, from left to right), and applied current $I(t)$ (bottom) from the data obtained from the sinusoidal current $I(t) = 10\sin(0.2t) + 10$ in \eqref{Eq:sine_curr}. In each panel, the EnKF estimated mean is shown in solid red while the true solution is shown in solid black. The dashed red lines show the estimated $\pm 2$ standard deviation curves around the mean.}
\label{Fig:sin_final}
\end{figure*}

To establish baseline results for parameter tracking in each of the four data sets described in Section~\ref{SubSec:Data}, the augmented EnKF was employed using $N=100$ ensemble members with the standard deviation of the parameter random walk in \eqref{Eq:rand_walk} set to $\sigma_\xi = 1$.  The results in Figures~\ref{Fig:constant_final}--\ref{Fig:sin_final} show the parameter tracking estimates of $I(t)$, along with the time series estimates of the Hodgkin-Huxley model states, for each of the four cases.  Note that in each case, the filter mean is able to track the underlying true applied current, along with the unmeasurable system states, with uncertainty bounds represented by the $\pm2$ estimated standard deviation curves.

As noted in Section~\ref{Sec:ParamEst}, a carefully chosen standard deviation $\sigma_\xi$ for the random walk in \eqref{Eq:rand_walk} is crucial in maintaining the accuracy of the time-varying parameter estimate and avoiding filter divergence; see \cite{Arnold2018} and references therein.  To demonstrate the sensitivity of the algorithm to this choice, we tested the effects of changing $\sigma_\xi$ by letting $\sigma_{\xi} = 10$, 2, 1, 0.5, 0.25, and 0.1 in estimating the applied current, compared with $\sigma_{\xi} = 1$ used in the baseline results above.  For sake of demonstration, we focused on the data generated using the sinusoidal current defined in \eqref{Eq:sine_curr}; similar results hold in the other cases.  

\begin{figure*}[h]
\centerline{\includegraphics[width=0.33\textwidth]{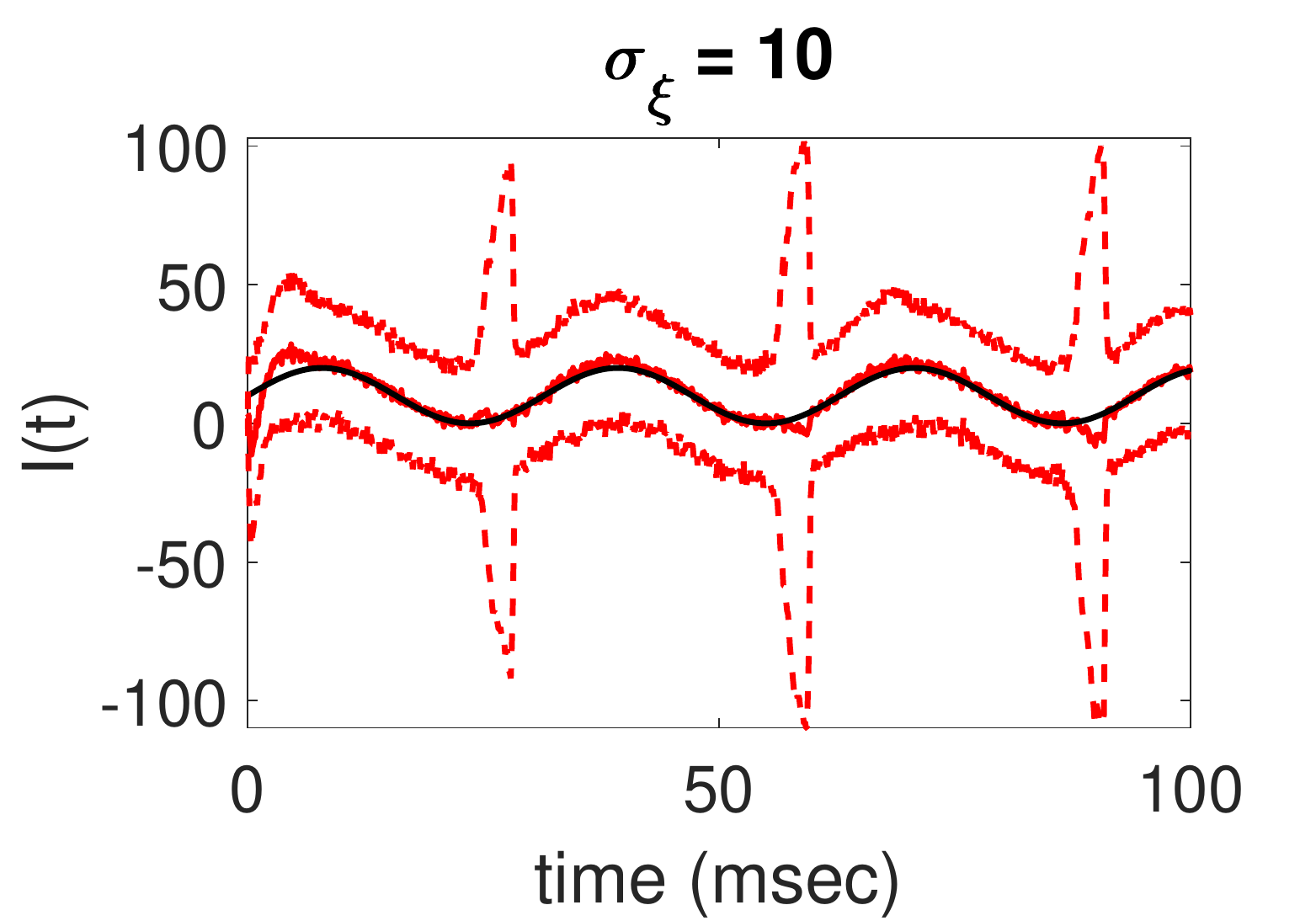} \includegraphics[width=0.33\textwidth]{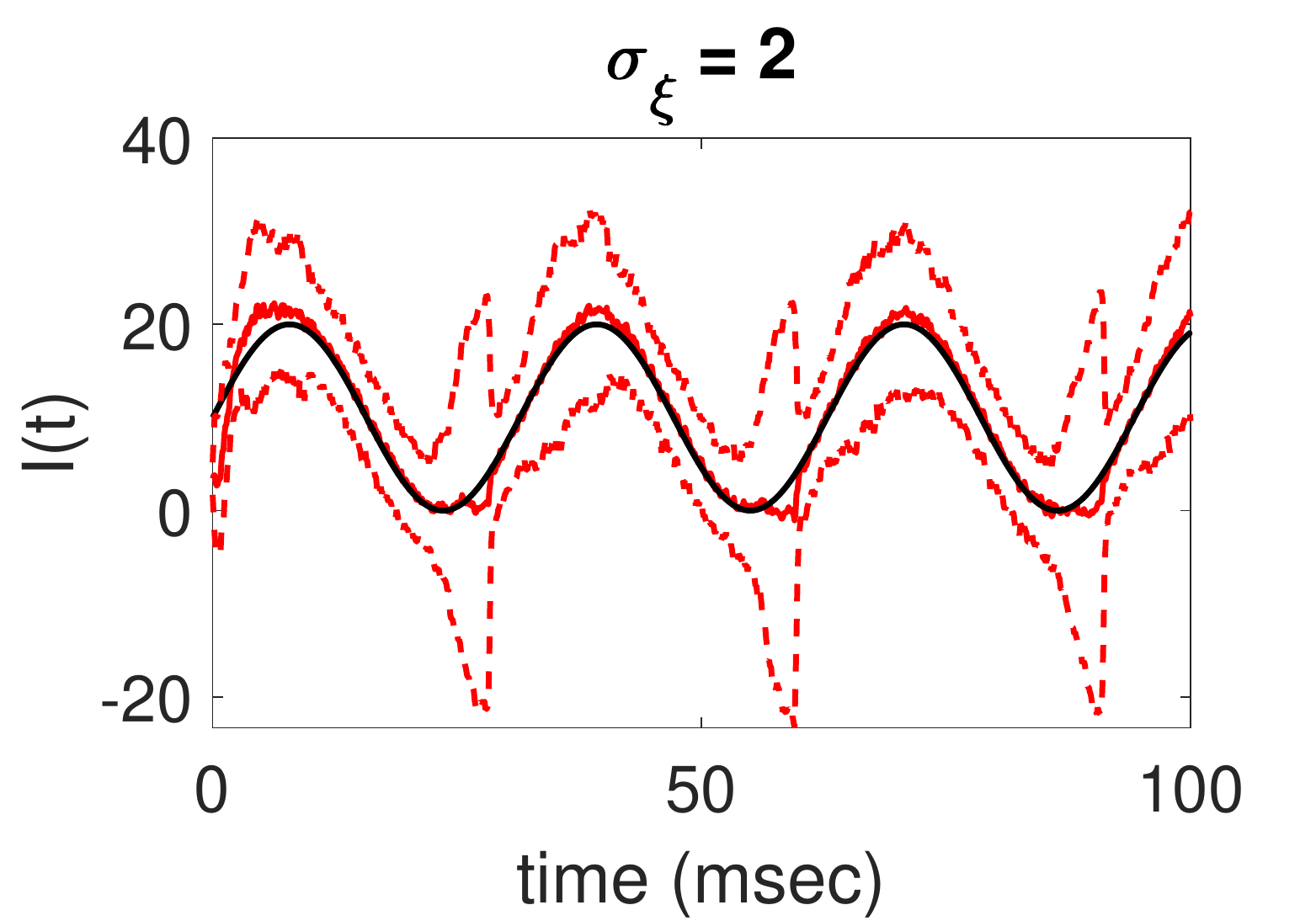} \includegraphics[width=0.33\textwidth]{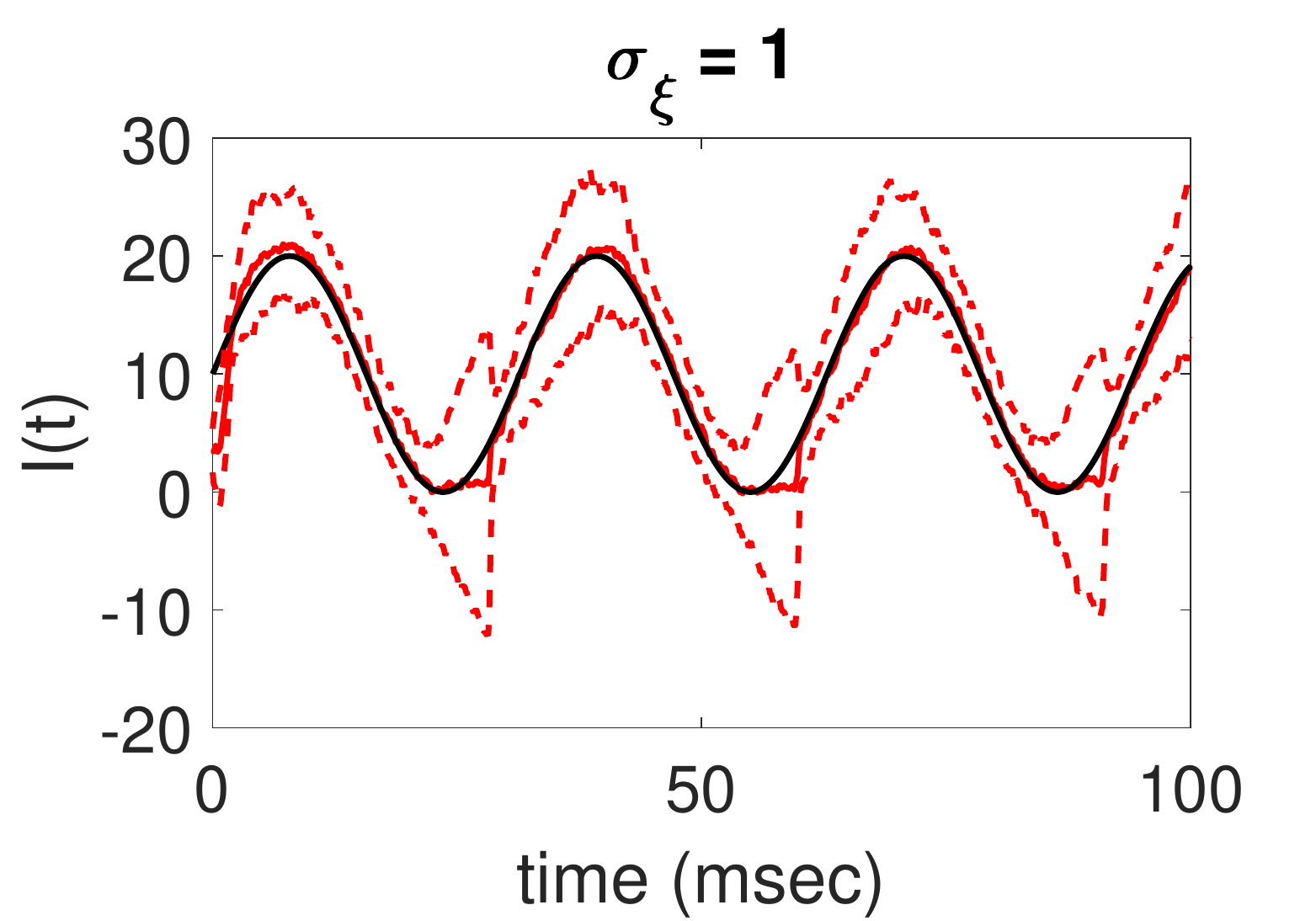} }
\vspace{.4cm}
\centerline{ \includegraphics[width=0.33\textwidth]{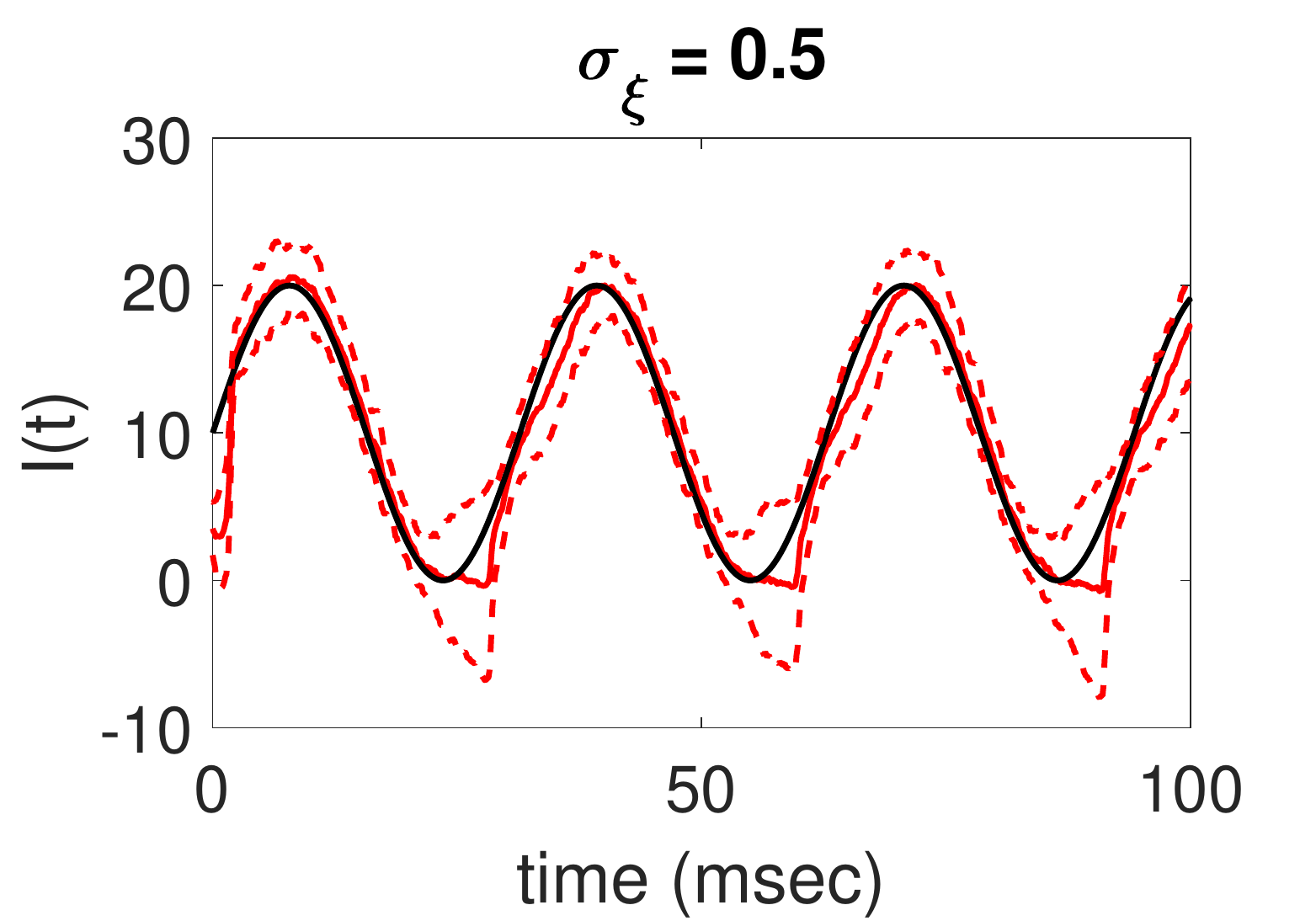} \includegraphics[width=0.33\textwidth]{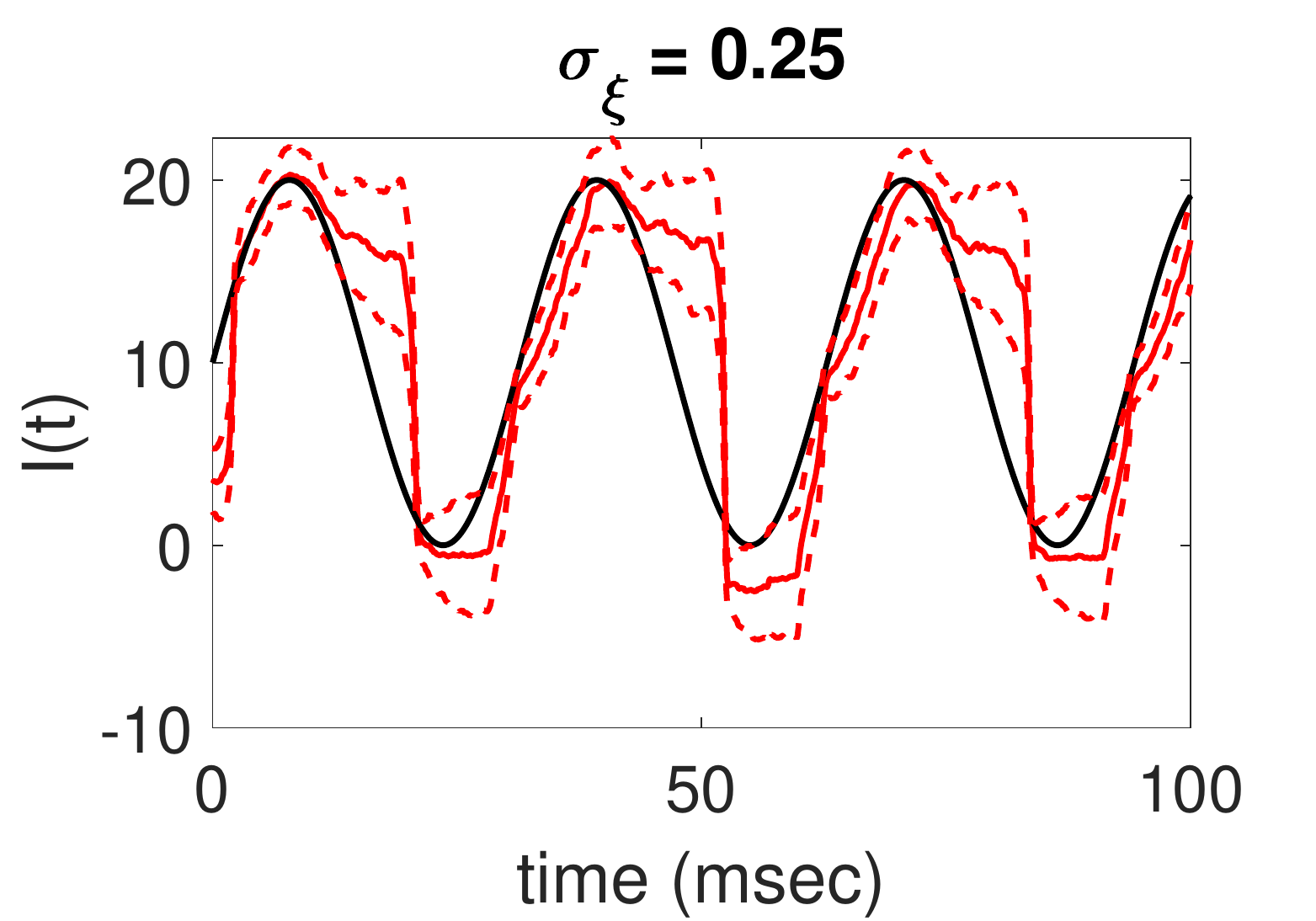}  \includegraphics[width=0.33\textwidth]{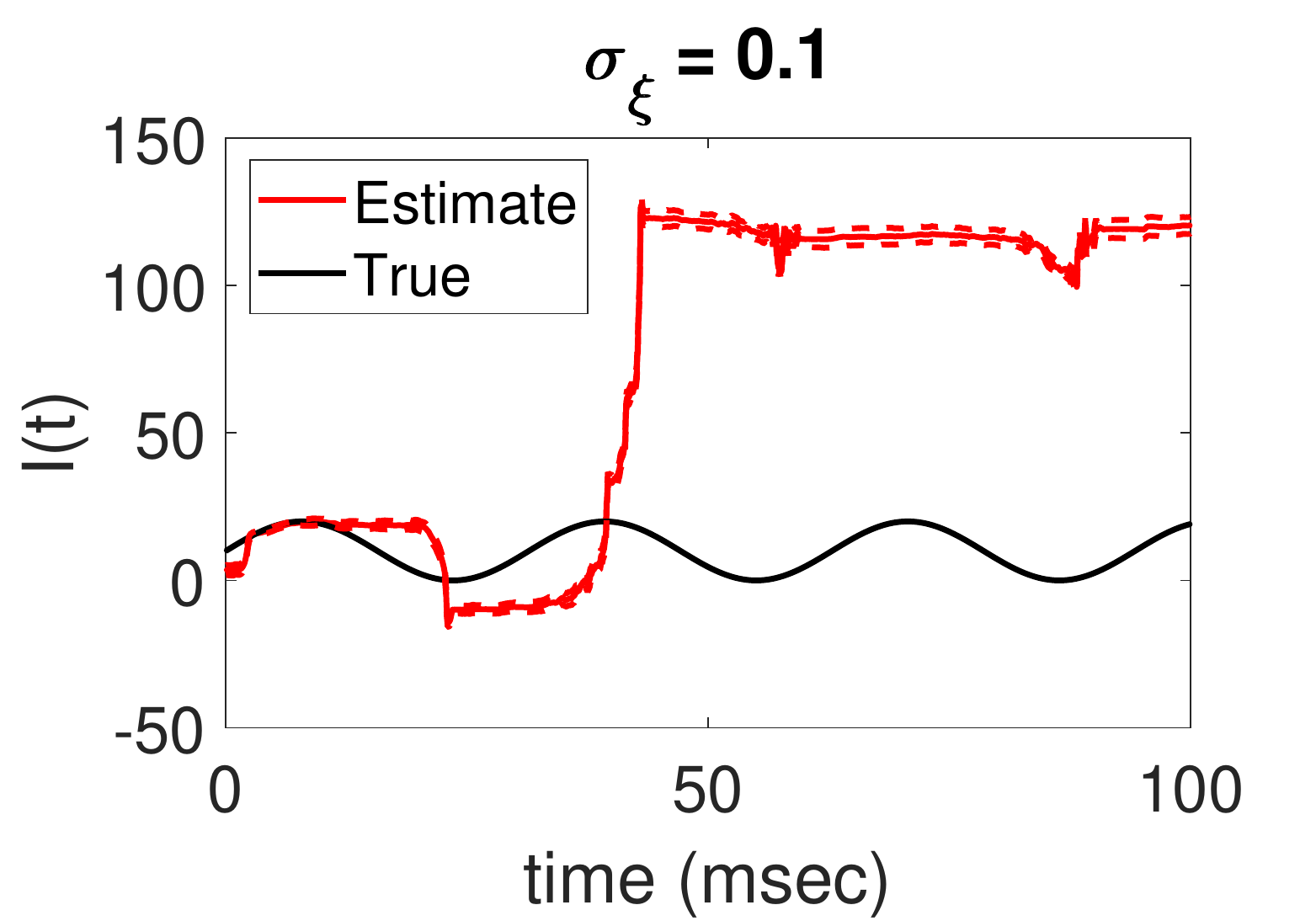} }
\caption{Resulting EnKF with parameter tracking estimates of $I(t)$ using different standard deviations $\sigma_{\xi}$ for the parameter tracking drift term in \eqref{Eq:rand_walk}.  In each panel, the solid black line is the true sinusoidal applied current $I(t) = 10\sin(0.2t) +10$ in \eqref{Eq:sine_curr} that the filter aims to estimate.  The solid red line is the EnKF estimated mean, and the dashed red lines show the estimated $\pm 2$ standard deviation curves around the mean.}
\label{Fig:Standard_devs}
\end{figure*}

The results in Figure \ref{Fig:Standard_devs} show that when $\sigma_{\xi} = 10$, although the EnKF mean estimate was able to well-track the true solution, the estimated $\pm2$ standard deviation curves around the mean are very large around the mean, reflecting a lack of confidence in the estimate.  On the other hand, when $\sigma_{\xi} = 0.1$, the filter is unable to well-track the true parameter and eventually diverges.  In this case, while the filter is unable to track the true parameter, the estimated $\pm2$ standard deviation curves are very tight around the mean, implying a high confidence in an incorrect estimate.  For this example, the choice of $\sigma_{\xi} = 0.5$ in the parameter random walk visually captures the true underlying parameter the best out of the values considered, as it keeps the EnKF estimated mean closely with a small standard deviation around the mean. 

In addition to the choice of $\sigma_\xi$, the frequency of time-series data available also has a significant effect on the resulting parameter tracking estimates.  To analyze this on the problem at hand, we subsampled the available voltage data corresponding to the pulsing step current \eqref{Eq:pulsing_curr} and sinusoidal current \eqref{Eq:sine_curr} every 10, 20, and 50 time points, resulting in data sets containing 201, 101, and 41 equidistant voltage observations over the time interval $[0,200]$.  This means that data was considered in the observation step in the EnKF every first, second, or fifth millisecond (as compared to every 0.1 msec using the full data).  Note that the filter still performed the prediction step every 0.1 msec as before, but the observation step was only preformed if data was available at that time point.  

\begin{figure*}[h!]
\centering{\includegraphics[width=0.75\textwidth]{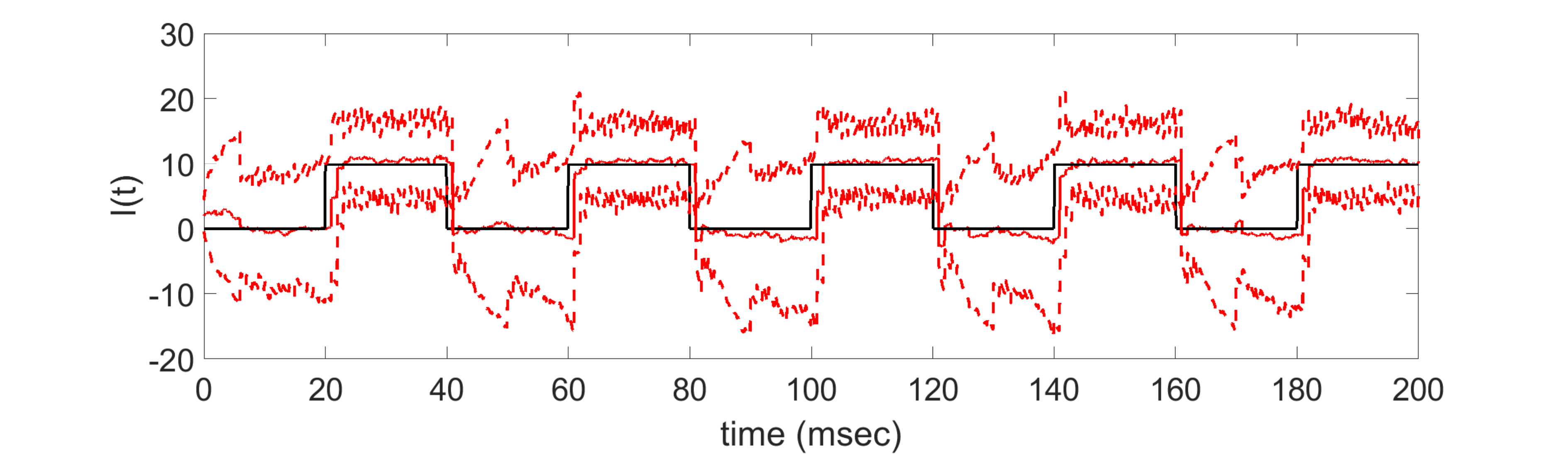}}
\vspace{.4cm}
\centering{\includegraphics[width=0.75\textwidth]{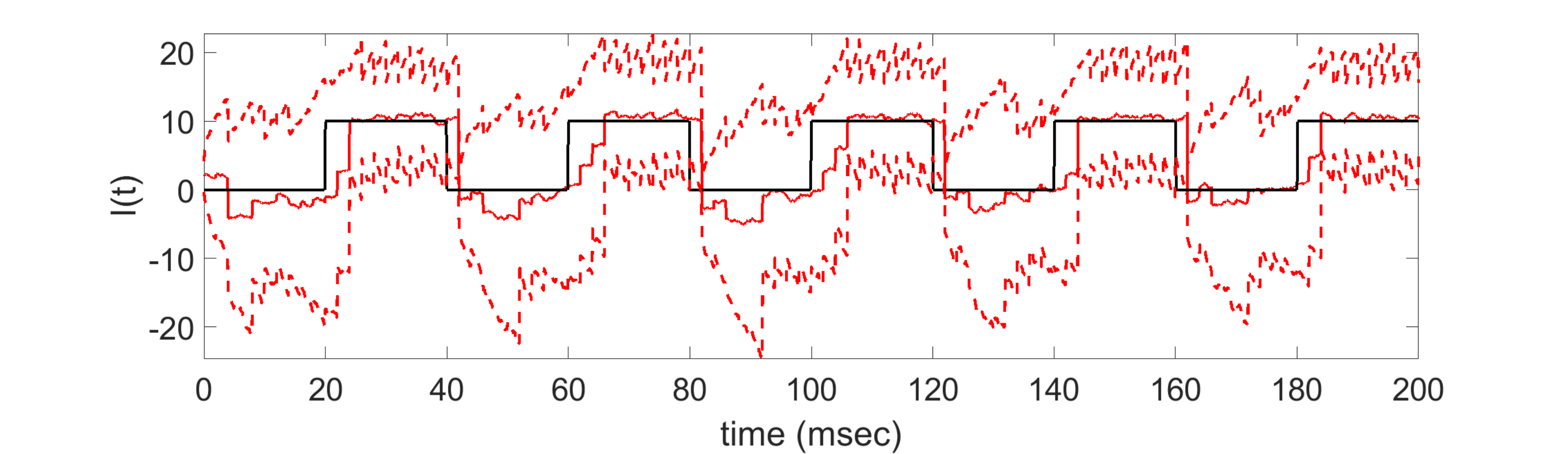}}
\vspace{.4cm}
\centering{\includegraphics[width=0.75\textwidth]{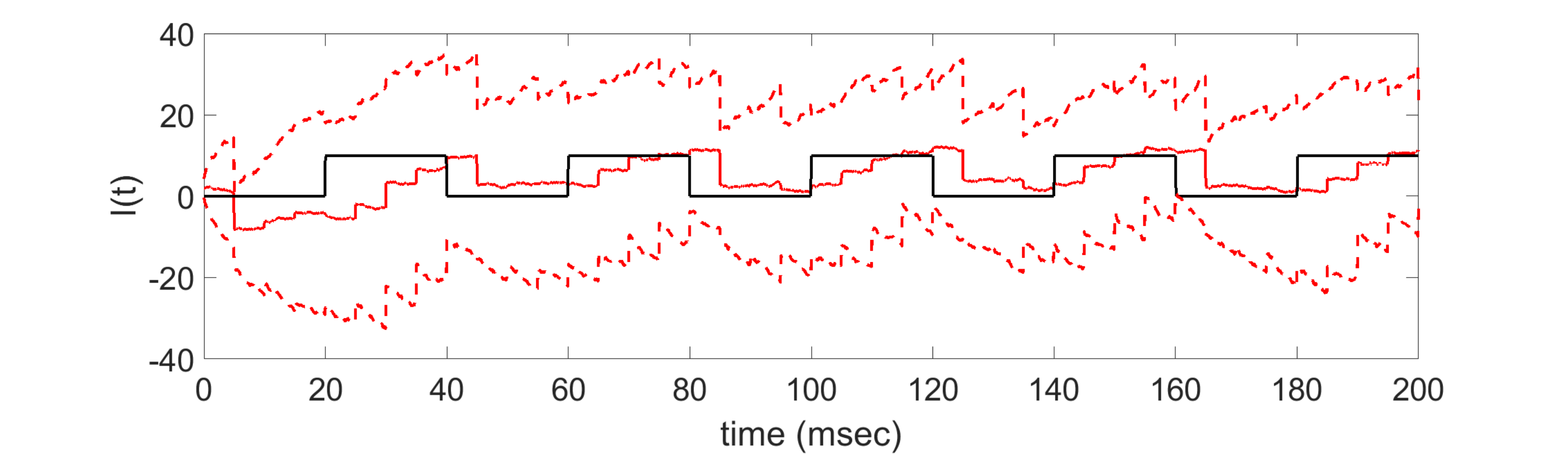}}
\caption{Resulting EnKF with parameter tracking estimates of $I(t)$ when the data generated using the pulsing step current in \eqref{Eq:pulsing_curr} is subsampled every 10 (top), 20 (middle), and 50 (bottom) time points.  In each panel, the solid black line shows the true applied current, the solid red line shows the EnKF estimated mean, and the dashed red lines show the estimated $\pm 2$ standard deviation curves around the mean.} 
\label{Fig:less_data_step}
\end{figure*}

\begin{figure*}[h!]
\centering{\includegraphics[width=0.75\textwidth]{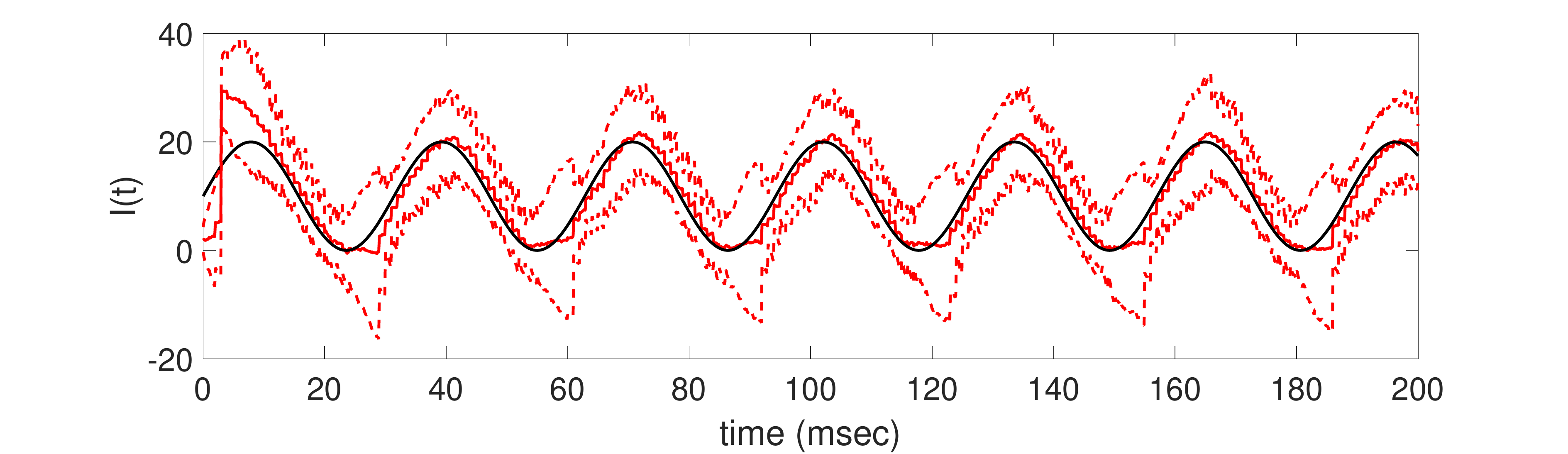}}
\vspace{.4cm}
\centering{\includegraphics[width=0.75\textwidth]{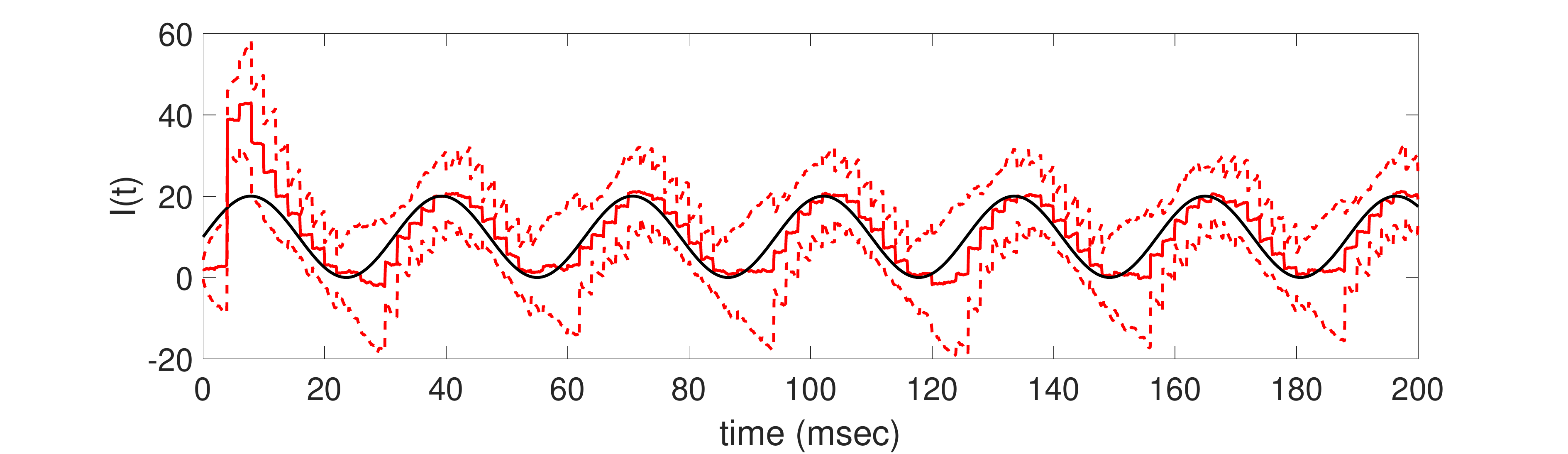}}
\vspace{.4cm}
\centering{\includegraphics[width=0.75\textwidth]{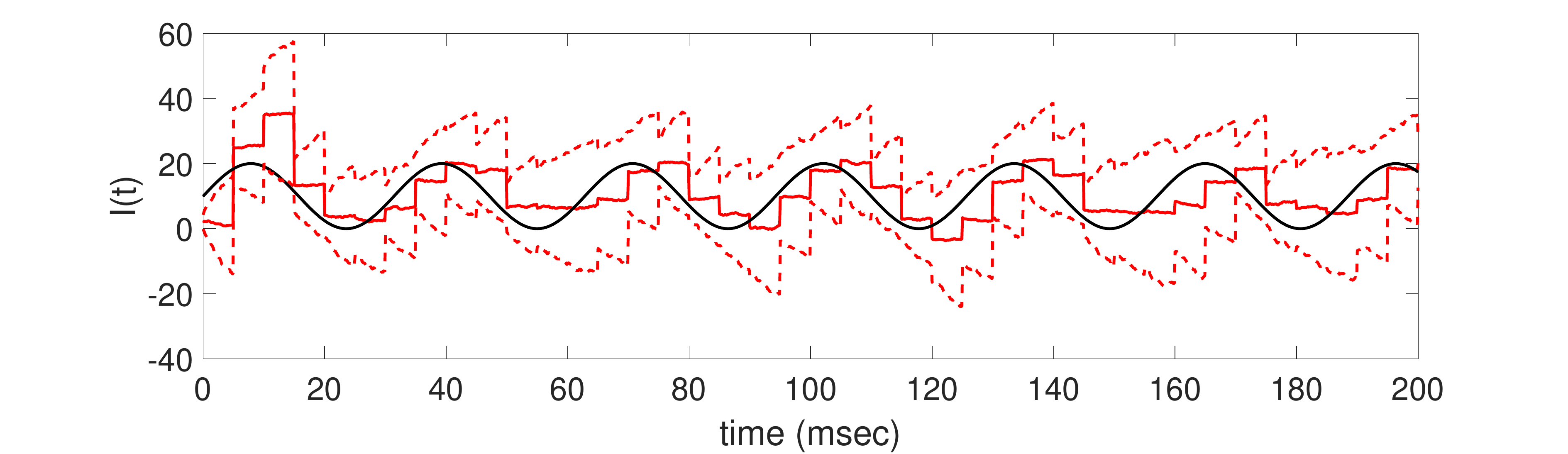}}
\caption{Resulting EnKF with parameter tracking estimates of $I(t)$ when the data generated using the sinusoidal current in \eqref{Eq:sine_curr} is subsampled every 10 (top), 20 (middle), and 50 (bottom) time points.  In each panel, the solid black line shows the true applied current, the solid red line shows the EnKF estimated mean, and the dashed red lines show the estimated $\pm 2$ standard deviation curves around the mean.}
\label{Fig:less_data_sin}
\end{figure*}

The results in Figures \ref{Fig:less_data_step} and \ref{Fig:less_data_sin} show that as data becomes more and more sparse, the parameter tracking estimate of the applied current parameter loses more and more characteristics of the underlying deterministic function.  In particular, in Figure \ref{Fig:less_data_step}, as less data is available, the parameter tracking estimate of the pulsing step function begins lagging in estimating the steps and is unable to well maintain the shape.  Similar results are seen in Figure \ref{Fig:less_data_sin} for the sinusoidal function, where the parameter tracking estimate loses its periodicity as less data becomes available.  In both figures, it is clear that the sparser the data set, the more challenging it becomes for the filter to track the true underlying applied current function.

\section{Discussion}

The aim of this work was to utilize the EnKF with parameter tracking in estimating the time-varying applied current parameter in the Hodgkin-Huxley model.  In particular, the parameter tracking algorithm was analyzed in estimating four different deterministic applied currents using synthetically-generated voltage data.  We first verified that the algorithm was able to successfully track the underlying applied current, along with the unobserved model states, for each of the four test cases.  In addition to tracking the applied current $I(t)$, baseline results show that the filter is able to accurately estimate the unobserved states $n(t)$, $m(t)$, and $h(t)$ from the generated voltage data.  Further numerical experiments were conducted to analyze how the parameter tracking estimates of the applied current parameter are affected under different implementation conditions, namely, when changing the standard deviation of the parameter drift term in \eqref{Eq:rand_walk} and when the algorithm is provided increasingly less amounts of data. 

In general, using the augmented EnKF with parameter tracking as described in Section~\ref{Sec:ParamEst}, we were able to well track the underlying applied current functions in each of the four cases considered, establishing the baseline results shown in Figures \ref{Fig:constant_final}--\ref{Fig:sin_final}.  From these results, we were able to further explore the effects of two different aspects of the parameter tracking implementation: the choice of the standard deviation of the parameter drift term in \eqref{Eq:rand_walk}, and the availability of time-series data.  

We found that using different values for $\sigma_{\xi}$ resulted in drastically different levels of accuracy and confidence in the resulting time-varying parameter estimates for the applied current.  As shown in Figure \ref{Fig:Standard_devs}, when using $\sigma_{\xi} = 10$ here for the parameter drift, the resulting mean estimate tracked well, however the resulting confidence in the estimate was low, as reflected in the wide range between the estimated $\pm 2$ standard deviation curves.  Oppositely, when using $\sigma_{\xi} = 0.1$, the parameter tracking estimate diverged, returning an estimate of the parameter which was far off from the true solution despite having high confidence in the estimate.  Based on the results in Figure \ref{Fig:Standard_devs}, it was determined that for the data considered, $\sigma_{\xi} = 0.5$ was the best choice out of the tested values, since the algorithm was able to accurately track the mean with less uncertainty reflected in the resulting $\pm2$ standard deviation curves around the mean.  

We also explored how the accuracy of the parameter tracking estimate changed when the generated data was subsampled, resulting in fewer data points being used as updating information in the filter.  The results in Figures \ref{Fig:less_data_step} and \ref{Fig:less_data_sin} show that as less data was made available, the parameter tracking algorithm had increasing difficulty in estimating the true underlying applied current function, losing structural features such as the step onset of the pulsing step current and the periodicity of the sinusoidal current.  

While the focus of this work was on estimating four different deterministic forms of the time-varying applied current parameter, in future work we aim to estimate stochastic forms of the current, as well as estimating applied currents relating to networks of neurons.  In addition to the applied current, we are also interested in applying parameter tracking methodology to estimate the $\alpha(V)$ and $\beta(V)$ rate functions as additional unknown parameters that vary with voltage, which would be particularly useful in applications of the Hodgkin-Huxley model where these rate functions do not necessarily share the same parameterized forms as in \eqref{Eq:m_alpha}--\eqref{Eq:h_beta} and \eqref{Eq:n_alpha}--\eqref{Eq:n_beta}.

Additional future work includes comparing our results for the Hodgkin-Huxley model with results using other single neuron models, such as the FitzHugh-Nagumo and Hindmarsh-Rose models \cite{FitzHugh1961, Nagumo1962, Hindmarsh1984}.  We also aim to apply our results to further biomedical applications utilizing Hodgkin-Huxley dynamics to model various neurodegenerative diseases affecting the function of neurons and ionic channels, such as Alzheimer's disease and amyotrophic lateral sclerosis \cite{Kagan2002, Kanai2006, Bostock1995}.


\vspace{6pt} 



\authorcontributions{conceptualization, A.A.; formal analysis, K.C., L.S. and A.A.; funding acquisition, A.A.; software, K.C., L.S. and A.A.; supervision, A.A.; validation, K.C., L.S. and A.A.; visualization, K.C., L.S. and A.A.; writing--original draft preparation, K.C., L.S. and A.A.; writing--review and editing, K.C., L.S. and A.A.}

\funding{This research was funded by the National Science Foundation grant number NSF/DMS-1819203.}


\conflictsofinterest{The authors declare no conflict of interest.} 

%

%

\reftitle{References}


\externalbibliography{yes}
\bibliography{Lit_References}{}






\end{document}